\documentclass[a4paper,usenatbib]{mn2e}
\usepackage{txfonts}
\pdfoutput=1
\usepackage{graphicx,ifthen,url,float,setspace}
\def\kms{km~s$^{-1}$ }

\def\s-1{s$^{-1}$}
\def\Hz-1{Hz$^{-1}$}

\def\l1l2{\lambda_2 \; {\rm and} \; \lambda_1}

\def\eht1{{\rm \hat e_1}}

\def\d3x{d^3 x}

\def\cm#1{\, {\rm cm^{#1}}}

\newcommand{\nhi}{$N_{\rm HI}$}

\newcommand{\dla}{DLA}
\newcommand{\dlas}{DLAs}
\newcommand{\htwo}{H$_{\rm 2}$}

\newcommand{\hi}{H\, I}

\newcommand{\ewsitwo}{$W_{\lambda 1526}$}
\newcommand{\ewciv}{$W_{\lambda 1548}$}

\newcommand{\fitcosmob}{$-$0.04 $\pm$ 0.13}

\newcommand{\linpdelvm}{0.55} %% linear pearson coeff of delvvm plot
\newcommand{\delvvma}{$-$3.61 $\pm$ 0.07}  %% a of delvvm
\newcommand{\delvvmb}{1.13 $\pm$ 0.03}  %% b of delvvm
\newcommand{\delvvmmed}{114}  %% median delv of entire sample
\newcommand{\delvvmahi}{$-$3.67 $\pm$ 0.10}  %% a of delvvm
\newcommand{\delvvmbhi}{1.19 $\pm$ 0.05}  %% b of delvvm
\newcommand{\delvvmalo}{$-$3.58 $\pm$ 0.09}  %% a of delvvm
\newcommand{\delvvmblo}{1.10 $\pm$ 0.04}  %% b of delvvm

\newcommand{\mediandelvhi}{125 $\pm$ 60 \kms }
\newcommand{\medianmtlhi}{$-$1.18 $\pm$ 0.49}
\newcommand{\mediandelvlo}{98 $\pm$ 71 \kms }
\newcommand{\medianmtllo}{$-$1.17 $\pm$ 0.37}
\newcommand{\delvzp}{0.75}

\newcommand{\sivma}{$-$0.71 $\pm$ 0.02}
\newcommand{\sivmb}{1.46 $\pm$ 0.03}
\newcommand{\sivmahi}{$-$0.65 $\pm$ 0.03}
\newcommand{\sivmbhi}{1.53 $\pm$ 0.05}
\newcommand{\sivmalo}{$-$0.76 $\pm$ 0.03}
\newcommand{\sivmblo}{1.40 $\pm$ 0.04}
\newcommand{\linpsi}{0.76} %% linear pearson coeff of SiIIEW plot

\newcommand{\ewhi}{0.49 $\pm$ 0.23}
\newcommand{\ewlo}{0.38 $\pm$ 0.22}

\newcommand{\ewpks}{0.77}

\newcommand{\hiz}{R12}
\newcommand{\zmedian}{2.74}  %% median zabs

\newcommand{\delvninty}{$\Delta v_{90}$}

\loadboldmathitalic 
\title [Magellan Uniform Survey of DLAs: Cosmic Metallicity Evolution]
{The Magellan Uniform Survey of Damped Lyman $\alpha$ Systems I:  Cosmic Metallicity Evolution\thanks{This paper includes data gathered with the 6.5 meter Magellan Telescopes located at Las Campanas Observatory, Chile.}}
\author[R.~A. Jorgenson et al.]
{
Regina~A. Jorgenson$^{1,2}$\thanks{NSF Astronomy and Astrophysics Postdoctoral Fellow; raj@ifa.hawaii.edu}, Michael~T. Murphy$^3$, Rodger Thompson$^4$\\
$^1$ Institute for Astronomy, University of Hawai'i, 2680 Woodlawn Dr, Honolulu, HI 96822, USA \\
$^2$ Institute of Astronomy, University of Cambridge, Madingley Road, Cambridge, CB3 0HA, UK \\ 
$^3$ Centre for Astrophysics and Supercomputing, Swinburne University of Technology, Hawthorn, Melbourne, VIC 3122, Australia \\ 
$^4$ Steward Observatory, University of Arizona, Tucson, AZ 85721, USA \\
}
\date{Submitted \today.}
\begin{document} 
\maketitle 

\begin{abstract}

We present the chemical abundance measurements of the first large, medium-resolution, uniformly selected damped Lyman-$\alpha$ system (\dla ) survey.  The sample contains 99 \dlas\ towards 89 quasars selected from the SDSS DR5 \dla\ sample in a uniform way. We analyze the metallicities and kinematic diagnostics, including the velocity width of 90\% of the optical depth, $\Delta v_{90}$, and the equivalent widths of the Si{\sc \,ii} $\lambda$1526 ($W_{\lambda1526}$), C{\sc \,iv} $\lambda$1548 and Mg{\sc \,ii} $\lambda$2796 transitions.  To avoid strong line-saturation effects on the metallicities measured in medium-resolution spectra (FWHM$\sim$71 km s$^{-1}$), we derived metallicities from metal transitions which absorbed at most 35\,\% of the quasar continuum flux.  We find the evolution in cosmic mean metallicity of the sample, $\langle Z \rangle$ = (\fitcosmob )$z-$(1.06$\pm$0.36), consistent with no evolution over the redshift range $z \sim [2.2, 4.4]$, but note that the majority of our sample falls at $z \sim [2.2, 3.5]$.  The apparent lack of metallicity evolution with redshift is also seen in a lack of evolution in the median \delvninty\ and \ewsitwo\ values.  While this result may  seem to conflict with other large surveys that have detected significant metallicity evolution, such as ~\cite{rafelski12} who found $\langle Z \rangle$ = ($-$0.22$\pm$0.03)$z -$(0.65$\pm$0.09) over $z \sim [0, 5]$, several tests show that these surveys are not inconsistent with our new result.  However, over the smaller redshift range covered by our uniformly-selected sample, the true evolution of the cosmic mean metallicity in DLAs may be somewhat flatter than the ~\cite{rafelski12} estimate.

\end{abstract}

\section{Introduction}\label{introduction}

The damped Lyman-$\alpha$ systems (\dlas ), quasar absorption line systems having neutral gas columns of \nhi\ $\geq\ 2 \times 10^{20}$ cm$^{-2}$, represent a unique laboratory for understanding the conversion of neutral gas into stars at high redshift.  Dominating the neutral gas mass density between z=[0, 5] ~\citep{wolfe05}, the \dlas\  are believed to host the reservoirs of neutral gas for star formation across cosmic time.  However, the exact connection between \dlas\ and high redshift star formation remains elusive.  

One of the most compelling pieces of evidence for a connection between \dlas\ and high redshift star formation is the evolution of the  cosmological mean metallicity (e.g. ~\cite{vladilo00, vladilo02, pro03met, kulkarni05, kulkarni07, kulkarni10, rafelski12}).  This quantity, denoted by $\langle Z \rangle$ and defined by ~\cite{lanzetta95} as $\Omega _{metals}$/$\Omega _{gas}$, describes the amount of metals contained in the gas of \dlas .  Given that \dlas\ trace the neutral gas mass density of the Universe over cosmic time and that they are believed to be the neutral gas reservoirs for star formation, a natural consequence of this picture implies that the cosmic metallicity of \dlas\ should increase with time as subsequent generations of supernovae enrich the gas.  Therefore, tracing this metallicity evolution to high redshifts can provide important constraints on models of galaxy formation and evolution ~\citep{dave11, fuma11, cen12}. 

%%F1
\begin{figure*}
%\plotone{Figures/h2histdouble.pdf}
\includegraphics[width=1.75\columnwidth]{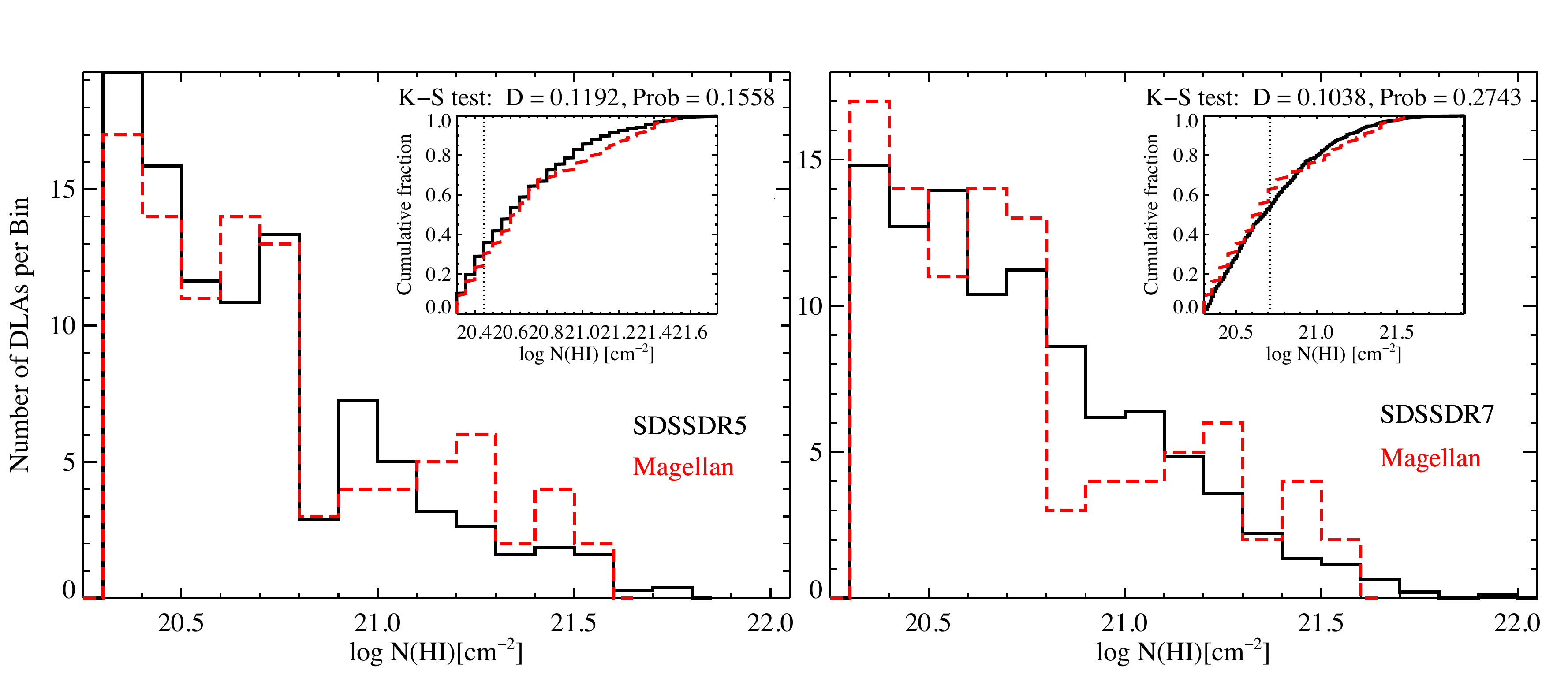}
\caption{\nhi\ histogram comparing the Magellan sample (red dashed line) with the scaled SDSS DR5 ~\citep{pro05} distribution (black), left, and with the scaled SDSS DR7 ~\citep{noterdaeme09} distribution, right.  A K-S test indicates the probability they are drawn from the same parent population is P$_{KS}$ = 0.16 and P$_{KS}$ = 0.27, respectively.
}
\label{fig:h2hist}
\end{figure*}

Contrary to expectations, initial studies (e. g. ~\cite{pettini99}) found no apparent evolution in the cosmic metallicity of \dla\ gas.  ~\cite{pettini99}  analyzed 40 \dlas\ in the redshift range $z~\sim$[0.5, 3.5] and reported no evidence for an increase of the column density-weighted metallicity below $z \sim 1.5$ as compared with higher redshifts.  They noted that this conflicts with the peak of the comoving star formation rate density that reaches a maximum between $z=1-2$.  ~\cite{pettini99} concluded that the apparent lack of evolution in metallicity could not be reconciled with the idea that \dlas\ fuelled the bulk of cosmic star formation at these epochs.   If, indeed, \dlas\ are an unbiased probe of this gas cycle, the cosmic mean metallicity of \dlas\ \emph{should} show a corresponding increase in step with the increase in the comoving star formation rate density of the Universe.

However, as samples grew in size, evolution in the cosmic mean metallicity of \dlas\ \emph{was} detected: first, by ~\cite{pro03met}, who found an evolution of $-0.26 \pm 0.07$ dex per unit redshift in a sample of 125 \dlas\ over the redshift range $z = [0.5, 5]$.  This evolution was confirmed in studies by ~\cite{kulkarni05, kulkarni07, kulkarni10}. Recently, the ~\cite{pro03met} sample was expanded by ~\cite{rafelski12}, who obtained medium and high-resolution follow-up of 30 high-$z$ ($z>4$) \dlas .  With an increased sample size of 242 \dlas , ~\cite{rafelski12} detected a 6$\sigma$ significant evolution in cosmic metallicity of $-0.22 \pm 0.03$ dex per unit redshift from $z = 0.09 - 5.06$.   
This apparent detection of evolution in metallicity over cosmic time may be important evidence linking \dla\ gas with star formation.  

With the aim of further elucidating the connection between \dlas\ and star formation, we recently completed a large, medium-resolution, uniformly selected survey of \dlas .  Our main motivation was to quantify the covering fraction of molecular hydrogen -- an important link between neutral gas and star formation -- so we took advantage of the exceptional blue throughput of the Magellan/MagE spectrograph in order to search for the redshifted UV Lyman and Werner band \htwo\ lines.  Our sample consists of 99 \dlas\ towards 89 quasars, selected from the SDSS DR5 catalog in a uniform way, i.e. including all \dlas\ visible from the Magellan site with $i$-band magnitude $\leq$19.  For convenience we will refer to this new sample of DLA spectra as the `Magellan sample' even though it comprises spectra from other telescopes as well (for some targets missed due to bad weather).
 Given the importance of understanding the metallicity evolution of \dlas\ and the effects of any potential biases in previous samples, we present in this paper the results of the metallicity analysis of the Magellan sample. We present the results of the search for \htwo\ in a second paper (Jorgenson et al. 2013b).%~\citep{jorgenson13b}.  

This paper is organized as follows.  We discuss our sample selection, observational details, and data reduction in \S~\ref{sec:data} while \S~\ref{sec:measure} contains a description of our procedure for measuring the metal-line column densities in each \dla .  In \S~\ref{sec:redshiftevol} we discuss the evidence for metallicity evolution.  An analysis of other \dla\ diagnostics is presented in \S~\ref{sec:others}.  Finally, we summarize the results and conclude in \S~\ref{sec:discussion}.

\section{Data \&\ Methodology}~\label{sec:data}

\subsection{Sample Selection}~\label{sec:sample}
In constructing the \dla\ sample presented in this paper, our primary goal was to determine the true covering factor and fraction of \htwo\ in \dlas .  To achieve this goal, we created a \dla\ sample drawn with uniform selection criteria from the SDSS DR5 DLA sample of ~\cite{pro05} with the aim of minimizing possible biases. We used just 3 simple selection criteria:  1) the target quasar had to be visible from the Magellan site (dec $\leq\ $ 15$^{\circ}$), 2) the redshift of the \dla\ was required to be $z_{abs}$ $\ge$ 2.2, such that the Lyman and Werner band molecular line region fell at $\lambda^{observed}$ $\ge$ 3200\AA\ and was observable from the ground and 3) the target quasar had an $i$-band magnitude of $i \leq\ $19, such that we created a reasonably sized sample that could be observed spectroscopically at moderate resolution with non-prohibitive amounts of telescope time.  This selection produced a total of 106 \dlas , towards 97 quasars.  

The resulting \dla\ sample, referred to here as the `Magellan sample,' is unique in the sense that it is the only large, uniformly selected \dla\ survey with medium-resolution (or higher) spectra allowing for metallicity measurements.  Because our sample was taken directly from the SDSS in an unbiased way, the \hi\ column density distribution, f(\nhi ), is fairly well matched to that of the SDSS DR5 survey (see Figure~\ref{fig:h2hist}, left).  A two sided Kolmogorov-Smirnov (K-S) test shows that the probability of our sample and the SDSS DR5 (left) and SDSS DR7 (right) ~\citep{noterdaeme09} sample to be drawn from the same parent population is P$_{KS}$ = 0.16, and P$_{KS}$ = 0.27, respectively.  While our sample was designed to be as unbiased as possible, we note that, like any other \dla\ sample created from the SDSS survey, our sample will contain any of the biases inherent in the SDSS sample.  While dust-bias of the magnitude-limited SDSS sample is likely not a major issue (see, for example, ~\cite{ellison01, murphy04, jorgenson06, vladilo08, frank2010, khare12}), it is more difficult to assess how other biases, such as those created by color selection (i.e. ~\cite{richards01, worseck11}), may affect the results.  For example, ~\cite{worseck11} showed that a certain population of quasars is systematically missed in the SDSS selection because of overlap with the stellar locus in color space.  While this may be an important issue, it is outside the scope of the current paper and we will not consider the implications of biases in the SDSS further.  

In contrast with other surveys that heavily relied on archival and previously published data to create samples used to measure the \htwo\ fraction ~\citep{ledoux06} or the cosmic mean metallicity ~\citep{rafelski12, pro03met} of \dlas , the sample presented here was created \emph{a priori} to be an \htwo -blind and independent representation of the \dla\ population without regard to N(HI), metallicity, kinematics, or any other property of the \dla\ system.  While we were unable to obtain spectra of 10 sample \dlas\ due to bad weather (discussed in detail in \S~\ref{sec:reduction}), these `missing' \dlas\ should not add any additional bias to the sample.  Therefore, we argue that our sample is a less biased sample than those contained in previously published surveys and as a result, represents an important check on the results of those inhomogeneously created samples.  We discuss the issue of potential sample biases in greater detail in \S~\ref{sec:bias}.

\subsection{Data Acquisition and Reduction}~\label{sec:reduction}

Spectra were observed primarily during four observing runs during December 2008, and January, June and July of 2009 with the Magellan Echellette (MagE) Spectrometer on the Magellan II Clay telescope at Las Campanas Observatory ~\citep{marshall08}.  MagE was chosen as the best instrument for this survey primarily because of its excellent blue sensitivity, required for observing the redshifted Lyman and Werner band molecular hydrogen transitions that fall in the restframe UV.  In addition, the moderate spectral resolution ($\sim$ 71 \kms ) and large continuous wavelength coverage of 3100 \AA\ to 1 micron allowed for a relatively large survey in a reasonable amount of time with excellent wavelength coverage of each \dla .  All of the MagE spectra were taken with a 1\arcsec .0 slit giving a FWHM resolution of $\approx$ 71 km s$^{-1}$.  Exposure times ranged from 1200 to 8100 seconds resulting in a median S/N in the optimally extracted spectra of S/N $\sim$30 per resolution element.  

Unfortunately, due to bad weather, 15 sample quasars were not observed with MagE.  Several of these `missed' targets were later obtained with the X-Shooter spectrograph ~\citep{dodorico04}.  In total 8 target \dlas\ towards 7 quasars were obtained with X-Shooter.  These spectra were observed with a 1\arcsec.0 slit giving a FWHM resolution of $\approx$ 59 km s$^{-1}$.  

In addition, if high resolution echelle spectra of a sample quasar already existed in the Keck/HIRES ~\citep{vogt94} and/or VLT/UVES ~\citep{dekker00} archives, we did not re-observe it with MagE or X-Shooter (though there are some exceptions).  Existing high-resolution spectra were obtained from the archives of the VLT/UVES and Keck/HIRES spectrographs for 21 and 8 \dlas , respectively. These spectra were typically observed with a slit producing a resolution of $\approx$ 7 km s$^{-1}$.  Details of the total exposure time and instrument used for each target are given in Table ~\ref{tab:sample}.  Note that if a target was observed with two different instruments it is listed in Table ~\ref{tab:sample} twice.  In these cases, we used the higher resolution spectra for subsequent analyses.  

In summary, the Magellan, X-Shooter and archival spectra comprise a dataset of 96 of the original 106 sample \dlas , towards 89 quasars.  We also include in the sample three additional \dlas\ that were discovered in the course of our observations, bringing our total \dla\ sample to 99.  See \S~\ref{sec:discoveries} for details of the newly discovered \dlas .  Details of the `missing' 10 \dlas\ are reported in Table~\ref{tab:missing}. We stress that these `missing' \dlas , as previously mentioned, were missed only because of bad weather during our July 2009 observing run -- note, that they all have RA$\approx$13 -- and not because of some selection bias against faint quasar magnitudes or other such property.  Therefore, while not ideal, their absence from our final sample should not induce any additional bias.  

%%F2
 \begin{figure}
\includegraphics[width=0.99\columnwidth]{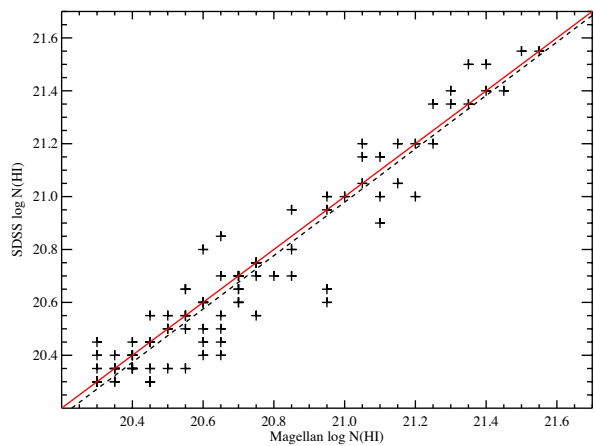}\\
\caption{
A comparison of the SDSS-derived \nhi\ measurements with those of the Magellan sample.  The red solid line indicates slope = 1, while the black dashed line is a least squares best-fit to the data with slope = 1.009.     
}
\label{fig:sdssnhi}
\end{figure}

 \begin{table*}
\centering
\caption{DLA Sample.}
{\scriptsize \begin{tabular}{ccccccccccccc}
\hline\hline
Quasar & $z_{em}$ & $z_{abs}$ & log N(HI) & $\Delta v$ & W$_{1526}$ & W$_{1548}$ & [M/H] & $f_M^a$ & [Fe/H]  & $f_{Fe}^b$ & I$^c$ & Exp. Time  \\
     &            &                &       [cm$^{-2}$]          &   [km s$^{-1}$]  &   [\AA ] & [\AA ] &     &                       &               &   &  & [s] \\
\hline\hline
J0011+1446&4.9672&3.4522&$21.40^{+0.20}_{-0.20}$& 425&$NA^d$&$NA^d$&$-1.12\pm0.20$& 1&$-1.70\pm0.11$& 4& 1&  1800\\
J0011+1446&4.9672&3.6175&$20.70^{+0.20}_{-0.20}$& 145&$NA^d$&$0.95\pm0.01$&$-2.47\pm0.18$&14&$-2.77\pm0.08$& 6& 1&  1800\\
J0013+1358&3.5755&3.2812&$21.50^{+0.10}_{-0.10}$&  65&$0.26\pm0.03$&$0.43\pm0.02$&$-1.81\pm0.17$& 1&$-2.14\pm0.23$&25& 1&  3000\\
J0035-0918&2.4195&2.3401&$20.50^{+0.10}_{-0.10}$&  25&$0.06\pm0.25$&$0.05\pm0.25$&$-2.61\pm0.12$& 1&$-2.37\pm0.09$& 1& 1&  4800\\
J0035-0918&2.4195&2.3401&$20.50^{+0.10}_{-0.10}$&  22&$0.04\pm0.01$&$0.02\pm0.01$&$-2.72\pm0.10$& 1&$-2.84\pm0.07$& 1& 4&  4800\\
J0124+0044&3.8292&3.0777&$20.30^{+0.10}_{-0.10}$& 142&$0.11\pm0.01$&$0.21\pm0.01$&$-1.59\pm0.27$&13&$-1.05\pm0.10$&11& 3& 19200$^{1,2,3}$\\
J0127+1405&2.4903&2.4416&$20.35^{+0.10}_{-0.10}$& 465&$1.13\pm0.02$&$0.77\pm0.02$&$-0.70\pm0.15$&13&$-1.37\pm0.04$& 1& 1&  1700\\
J0139-0824&3.0162&2.6773&$20.70^{+0.15}_{-0.15}$& 108&$0.61\pm0.01$&$0.35\pm0.01$&$-1.23\pm0.20$& 1&$-1.07\pm0.05$& 4& 3&  4800$^{4}$\\
J0211+1241&2.9531&2.5947&$20.60^{+0.10}_{-0.10}$& 485&$1.06\pm0.03$&$2.24\pm0.03$&$-0.58\pm0.12$& 1&$-0.47\pm0.02$& 4& 1&  1800\\
J0234-0751&2.5276&2.3181&$20.85^{+0.10}_{-0.10}$&  45&$0.13\pm0.01$&$0.15\pm0.03$&$-2.46\pm0.10$& 1&$-2.56\pm0.04$& 1& 1&  6000\\
J0239-0038&3.0751&3.0185&$20.35^{+0.10}_{-0.10}$& 185&$0.62\pm0.02$&$1.17\pm0.02$&$-0.46\pm0.13$& 1&$-0.51\pm0.07$& 4& 1&  1800\\
J0255+0048&3.9889&3.2540&$20.70^{+0.10}_{-0.10}$& 205&$1.08\pm0.02$&$0.91\pm0.02$&$-0.53\pm0.10$& 1&$-0.84\pm0.06$& 4& 1&  3000\\
J0255+0048&3.9889&3.9146&$21.30^{+0.10}_{-0.10}$&  45&$0.26\pm0.01$&$3.14\pm0.02$&$-1.70\pm0.10$& 4&$-1.78\pm0.07$& 4& 1&  3000\\
J0338-0005&3.0500&2.2297&$21.10^{+0.10}_{-0.10}$& 165&$1.32\pm0.44$&$0.52\pm0.44$&$-1.34\pm0.13$& 1&$-1.21\pm0.11$& 1& 1&  3600\\
J0338-0005&3.0500&2.2297&$21.10^{+0.10}_{-0.10}$& 227&$1.12\pm0.01$&$0.49\pm0.01$&$-1.22\pm0.11$& 1&$-1.74\pm0.10$&11& 3& 11800$^{4,5}$\\
J0912+0547&3.2406&3.1236&$20.30^{+0.10}_{-0.10}$&  45&$NA^d$&$0.05\pm0.02$&$-1.53\pm0.17$&14&$-1.83\pm0.06$& 1& 1&  1500\\
J0927+0746&2.5396&2.3104&$20.80^{+0.10}_{-0.10}$& 225&$0.49\pm0.01$&$0.55\pm0.01$&$-1.12\pm0.12$& 1&$-2.13\pm0.03$& 1& 1&  6300\\
J0942+0422&3.2755&2.3067&$20.30^{+0.20}_{-0.20}$& 138&$0.32\pm0.01$&$0.48\pm0.01$&$-0.80\pm0.20$& 1&$-0.98\pm0.02$& 4& 4&  7200\\
J0949+1115&3.8237&2.7584&$20.95^{+0.10}_{-0.10}$&  25&$NA^d$&$1.40\pm0.01$&$-1.18\pm0.12$& 1&$-1.14\pm0.07$& 1& 1&  2700\\
J0954+0915&3.3795&2.4420&$21.15^{+0.10}_{-0.10}$& 165&$1.27\pm0.01$&$NA^d$&$-1.16\pm0.11$& 1&$-1.49\pm0.10$& 1& 1&  2400\\
J1004+0018&3.0448&2.6855&$21.25^{+0.10}_{-0.10}$&  65&$0.21\pm0.01$&$0.30\pm0.02$&$-1.59\pm0.12$& 1&$-1.73\pm0.06$& 4& 1&  2400\\
J1004+0018&3.0448&2.5400&$21.10^{+0.10}_{-0.10}$& 125&$0.58\pm0.01$&$0.40\pm0.02$&$-1.12\pm0.11$& 1&$-1.23\pm0.03$& 4& 1&  2400\\
J1004+1202&2.8550&2.7997&$20.95^{+0.10}_{-0.10}$& 305&$1.55\pm0.01$&$1.60\pm0.01$&$-0.25\pm0.10$& 2&$-0.76\pm0.02$& 1& 1&  8100\\
J1019+0825&3.0104&2.3158&$20.30^{+0.10}_{-0.10}$&  45&$0.09\pm0.01$&$0.44\pm0.01$&$-2.21\pm0.11$& 1&$-2.53\pm0.03$& 1& 1&  1500\\
J1020+0922&3.6433&2.5931&$21.45^{+0.10}_{-0.10}$&  45&$NA^d$&$NA^d$&$-1.71\pm0.11$& 1&$-1.48\pm0.06$& 1& 1&  3400\\
J1022+0443&3.0811&2.7416&$20.60^{+0.10}_{-0.10}$& 185&$1.10\pm0.02$&$1.63\pm0.02$&$-0.83\pm0.13$& 1&$-0.72\pm0.04$& 4& 1&  2600\\
J1023+0709&3.7947&3.3777&$20.40^{+0.10}_{-0.10}$& 125&$0.14\pm0.02$&$0.17\pm0.01$&$-2.17\pm0.11$& 1&$-2.14\pm0.13$& 1& 1&  3200\\
J1029+1356&3.1159&2.9938&$20.70^{+0.10}_{-0.10}$& 105&$0.26\pm0.02$&$0.05\pm0.02$&$-1.33\pm0.15$&13&$-1.38\pm0.00$& 3& 1&  2700\\
J1032+0149&2.4275&2.2100&$20.45^{+0.10}_{-0.10}$&  85&$0.32\pm0.01$&$0.43\pm0.02$&$-0.88\pm0.13$& 1&$-1.81\pm0.03$& 1& 1&  6300\\
J1037+0910&3.5784&2.8426&$21.05^{+0.10}_{-0.10}$& 125&$0.36\pm0.01$&$0.21\pm0.02$&$-1.48\pm0.15$&13&$-1.55\pm0.12$& 1& 1&  2700\\
J1040-0015&4.2994&3.5450&$20.75^{+0.10}_{-0.10}$& 105&$0.58\pm0.01$&$0.34\pm0.02$&$-1.13\pm0.14$& 1&$-0.85\pm0.04$& 4& 1&  3000\\
J1042+0117&2.4407&2.2667&$20.75^{+0.10}_{-0.10}$& 118&$0.93\pm0.01$&$0.73\pm0.01$&$-0.68\pm0.10$& 1&$-1.09\pm0.09$& 1& 3&  4800$^{6}$\\
J1048+1331&3.1055&2.9196&$20.65^{+0.10}_{-0.10}$& 205&$0.65\pm0.02$&$0.50\pm0.02$&$-0.78\pm0.12$& 1&$-0.80\pm0.11$& 1& 1&  2700\\
J1057+0629&3.1423&2.4995&$20.50^{+0.10}_{-0.10}$& 405&$1.65\pm0.55$&$1.55\pm0.55$&$-0.36\pm0.11$& 1&$-0.76\pm0.13$& 1& 1&  2000\\
J1057+0629&3.1423&2.4995&$20.50^{+0.10}_{-0.10}$& 253&$1.57\pm0.01$&$1.49\pm0.01$&$-0.41\pm0.10$& 1&$-0.71\pm0.05$& 4& 3&  3600$^{7}$\\
J1100+1122&4.7068&4.3949&$21.40^{+0.20}_{-0.20}$& 128&$0.63\pm0.01$&$0.12\pm0.01$&$-1.46\pm0.16$&14&$-1.76\pm0.02$& 4& 2&  5181\\
J1100+1122&4.7068&3.7559&$20.75^{+0.20}_{-0.20}$&  93&$0.49\pm0.01$&$1.63\pm0.01$&$-1.42\pm0.21$& 1&$-0.92\pm0.00$& 3& 2&  5181\\
J1106+0816&4.2670&3.2240&$20.45^{+0.10}_{-0.10}$&  85&$0.22\pm0.01$&$0.27\pm0.01$&$-1.98\pm0.20$&14&$-2.28\pm0.13$& 1& 1&  2700\\
J1108+1209&3.6716&3.3963&$20.65^{+0.10}_{-0.10}$&  45&$0.09\pm0.01$&$0.19\pm0.01$&$-2.29\pm0.10$& 1&$-2.69\pm0.05$& 1& 3&  4200$^{6}$\\
J1108+1209&3.6716&3.5454&$20.75^{+0.10}_{-0.10}$& 128&$0.75\pm0.01$&$0.35\pm0.01$&$-1.05\pm0.11$& 1&$-1.40\pm0.02$& 4& 3&  4200$^{6}$\\
J1111+0714&2.8906&2.6820&$20.60^{+0.10}_{-0.10}$&  65&$0.23\pm0.01$&$1.18\pm0.01$&$-1.32\pm0.15$&13&$-2.06\pm0.06$& 1& 1&  2400\\
J1111+1332&2.4195&2.2710&$20.50^{+0.10}_{-0.10}$&  25&$0.07\pm0.01$&$0.10\pm0.01$&$-2.49\pm0.10$& 1&$-3.01\pm0.03$& 1& 1&  1800\\
J1111+1332&2.4195&2.3822&$20.45^{+0.10}_{-0.10}$&  25&$0.29\pm0.01$&$0.33\pm0.01$&$-1.27\pm0.15$&13&$-1.81\pm0.02$& 1& 1&  1800\\
J1111+1336&3.4816&3.2004&$21.15^{+0.10}_{-0.10}$& 153&$0.69\pm0.01$&$0.06\pm0.01$&$-2.18\pm0.15$&13&$-1.50\pm0.07$& 1& 3&  3600$^{6}$\\
J1111+1442&3.0916&2.5996&$21.35^{+0.15}_{-0.15}$& 190&$0.71\pm0.01$&$0.11\pm0.01$&$-1.19\pm0.16$&14&$-1.49\pm0.03$& 4& 3&  4500$^{7}$\\
J1133+0224&3.9899&3.9155&$20.55^{+0.10}_{-0.10}$&  25&$0.15\pm0.01$&$NA^d$&$-1.66\pm0.17$&14&$-1.97\pm0.07$& 1& 1&  2700\\
J1133+1305&3.6589&2.5975&$20.55^{+0.10}_{-0.10}$&  40&$0.52\pm0.01$&$0.22\pm0.01$&$-1.17\pm0.15$&13&$-1.94\pm0.09$& 1& 1&  1200\\
J1140+0546&3.0197&2.8847&$20.35^{+0.10}_{-0.10}$& 125&$0.27\pm0.02$&$0.63\pm0.02$&$-0.95\pm0.15$&13&$-0.50\pm0.04$& 4& 1&  2400\\
J1142-0012&2.4858&2.2578&$20.40^{+0.10}_{-0.10}$&  25&$0.31\pm0.01$&$0.20\pm0.01$&$-0.85\pm0.17$&13&$-1.64\pm0.02$& 1& 1&  4200\\
J1151+0552&3.2406&2.9287&$20.85^{+0.10}_{-0.10}$& 165&$1.32\pm0.01$&$0.70\pm0.02$&$-1.03\pm0.12$& 1&$-1.05\pm0.07$& 1& 1&  3000\\
J1153+1011&4.1272&3.7950&$21.35^{+0.10}_{-0.10}$&  25&$0.26\pm0.02$&$0.45\pm0.02$&$-2.07\pm0.15$&13&$-2.08\pm0.02$& 2& 1&  2700\\
J1153+1011&4.1272&3.4695&$20.75^{+0.10}_{-0.10}$&  65&$0.30\pm0.01$&$0.92\pm0.02$&$-0.96\pm0.14$& 1&$-1.52\pm0.00$& 3& 1&  2700\\
J1155+0530&3.4752&3.3261&$21.05^{+0.10}_{-0.10}$& 213&$1.17\pm0.01$&$0.28\pm0.01$&$-0.65\pm0.10$& 1&$-1.32\pm0.03$& 1& 3& 18600$^{9,6}$\\
J1155+0530&3.4752&2.6077&$20.50^{+0.10}_{-0.10}$&  27&$0.11\pm0.01$&$0.14\pm0.01$&$-1.76\pm0.14$& 1&$-2.10\pm0.01$& 1& 3& 18600$^{9,6}$\\
J1201+0116&3.2330&2.6852&$21.00^{+0.15}_{-0.15}$&  91&$0.41\pm0.01$&$0.38\pm0.01$&$-1.77\pm0.16$&14&$-2.07\pm0.01$& 1& 4&  4775\\
J1208+0043&2.7213&2.6084&$20.45^{+0.10}_{-0.10}$& 205&$0.24\pm0.01$&$0.28\pm0.01$&$-1.92\pm0.10$& 1&$-1.04\pm0.12$& 4& 1&  2700\\
J1211+0422&2.5416&2.3766&$20.65^{+0.10}_{-0.10}$& 114&$0.29\pm0.01$&$0.90\pm0.01$&$-1.19\pm0.11$& 1&$-1.58\pm0.05$& 4& 4&  9000\\
J1211+0902&3.2905&2.5835&$21.30^{+0.10}_{-0.10}$& 328&$1.62\pm0.01$&$1.78\pm0.01$&$-0.84\pm0.10$& 1&$-1.09\pm0.01$& 4& 3& 27365$^{10,11,7}$\\
J1220+0921&4.1103&3.3090&$20.40^{+0.20}_{-0.20}$& 125&$0.07\pm0.01$&$1.65\pm0.01$&$-2.48\pm0.22$& 1&$-1.89\pm0.00$& 3& 1&  3000\\
J1223+1034&2.7613&2.7194&$20.45^{+0.10}_{-0.10}$& 345&$0.45\pm0.02$&$2.85\pm0.02$&$-1.14\pm0.15$&13&$-1.88\pm0.07$& 1& 1&  2100\\
J1226+0325&2.9769&2.5078&$20.95^{+0.10}_{-0.10}$& 445&$1.03\pm0.02$&$1.11\pm0.02$&$-0.96\pm0.11$& 1&$-1.78\pm0.04$& 1& 1&  2400\\
J1226-0054&2.6169&2.2903&$20.70^{+0.10}_{-0.10}$& 305&$1.84\pm0.01$&$1.23\pm0.02$&$-0.88\pm0.13$& 1&$-0.83\pm0.05$& 4& 1&  3000\\
J1228-0104&2.6553&2.2625&$20.40^{+0.10}_{-0.10}$&  98&$0.64\pm0.43$&$0.52\pm0.53$&$-0.92\pm0.12$& 1&$-1.41\pm0.01$& 1& 3&  3600$^{7}$\\
\hline\hline
\end{tabular}
}

\label{tab:sample}
\begin{spacing}{0.7}
{\scriptsize {\bf $^a$} Flag describing the metallicity measurement: (1) Si measurement; (2) Zn measurement; (4) = S measurement, (13) mix of limits; (14) Fe measurement + 0.3; (15) Fe Limit + 0.3  } \\
{\scriptsize {\bf $^b$} Flag describing the Fe Measurement: (1) Fe abundance; (2) Fe lower limit; (3) Fe upper limit; (4) Ni abundance offset by -0.1; (5) Cr abundance offset by -0.2; (6) Al abundance; (11) Fe limits from a pair of transitions; (13) Limit from Fe+Ni  } \\
{\scriptsize {\bf $^c$} Instrument Used: 1 = MagE (FWHM$\sim$71 km s$^{-1}$), 2 = XShooter (FWHM$\sim$59 km s$^{-1}$), 3 = UVES (FWHM$\sim$8 km s$^{-1}$), 4 = HIRES (FWHM$\sim$8 km s$^{-1}$) } \\
{\scriptsize {\bf $^d$} NA indicates either no spectral coverage or severe blending that precluded an equivalent width measurement} \\
{\scriptsize {\bf Table Notes.} 
--VLT Program ID Number: 
1: 069.A-0613;
2: 071.A-0114;
3: 073.A-0653;
4: 074.A-0201;
5: 080.A-0014;
6: 080.A-0482;
7: 081.A-0334;
8: 080.A-0482;
9: 076.A-0376;
10: 067.A-0146;
11: 073.B-0787;
12: 067.A-0078;
13: 068.A-0600;
14: 072.A-0346;
15: 079.A-0404
} 
\end{spacing}
\end{table*}

 \begin{table*}
\contcaption{DLA Sample.}
\centering
%\caption{DLA Sample.}
{\scriptsize \begin{tabular}{ccccccccccccc}
\hline\hline
Quasar & $z_{em}$ & $z_{abs}$ & log N(HI) & $\Delta v$ & W$_{1526}$ & W$_{1548}$ & [M/H] & $f_M^a$ & [Fe/H]  & $f_{Fe}^b$ & I$^c$ & Exp. Time  \\
     &            &                &       [cm$^{-2}$]          &   [km s$^{-1}$]  &   [\AA ] & [\AA ] &     &                       &               &   &  & [s] \\
\hline\hline
J1228-0104&2.6553&2.2625&$20.40^{+0.10}_{-0.10}$& 125&$0.44\pm0.01$&$0.24\pm0.01$&$-0.90\pm0.16$&13&$-0.57\pm0.08$& 4& 1&  2400\\
J1233+1100&2.8857&2.8206&$20.35^{+0.10}_{-0.10}$& 185&$0.69\pm0.03$&$0.49\pm0.03$&$-0.54\pm0.23$&13&$-1.52\pm0.05$& 1& 1&  3000\\
J1233+1100&2.8857&2.7924&$20.60^{+0.10}_{-0.10}$& 365&$1.02\pm0.03$&$0.90\pm0.03$&$-0.48\pm0.11$& 1&$-0.69\pm0.14$& 1& 1&  3000\\
J1240+1455&3.0847&3.0242&$20.30^{+0.10}_{-0.10}$& 117&$0.37\pm0.01$&$0.44\pm0.01$&$-1.52\pm0.10$& 1&$-1.49\pm0.02$& 1& 3& 26881$^{5}$\\
J1246+1113&3.1541&3.0971&$20.45^{+0.10}_{-0.10}$& 165&$1.00\pm0.02$&$1.08\pm0.02$&$-0.53\pm0.14$& 1&$-0.55\pm0.04$& 4& 1&  1200\\
J1253+1147&3.2851&2.9443&$20.40^{+0.10}_{-0.10}$&  75&$0.44\pm0.01$&$0.29\pm0.01$&$-0.79\pm0.11$& 1&$-1.12\pm0.09$& 4& 3&  3600$^{7}$\\
J1253+1306&3.6244&2.9812&$20.60^{+0.10}_{-0.10}$&  65&$0.48\pm0.01$&$0.18\pm0.01$&$-1.07\pm0.18$&13&$-1.14\pm0.11$& 4& 1&  2700\\
J1257-0111&4.1117&4.0209&$20.35^{+0.10}_{-0.10}$& 225&$0.58\pm0.01$&$0.87\pm0.01$&$-0.88\pm0.10$& 4&$-1.54\pm0.05$& 1& 1&  2100\\
J1304+1202&2.9805&2.9133&$20.55^{+0.10}_{-0.10}$& 125&$0.30\pm0.02$&$0.82\pm0.02$&$-1.15\pm0.11$& 4&$-1.89\pm0.06$& 1& 1&  1200\\
J1304+1202&2.9805&2.9288&$20.35^{+0.10}_{-0.10}$&  45&$0.16\pm0.02$&$0.18\pm0.01$&$-1.96\pm0.11$& 1&$-2.28\pm0.09$& 1& 1&  1200\\
J1306-0135&2.9422&2.7730&$20.60^{+0.20}_{-0.20}$& 165&$0.77\pm0.08$&$0.35\pm0.08$&$-0.69\pm0.18$&13&$-0.07\pm0.10$&11& 1&  1000\\
J1309+0254&2.9392&2.2450&$20.70^{+0.10}_{-0.10}$& 192&$0.89\pm0.02$&$0.45\pm0.03$&$-0.24\pm0.15$&13&$-0.89\pm0.10$&11& 2&  2400\\
J1317+0100&2.6984&2.5365&$21.55^{+0.15}_{-0.15}$&  12&$0.37\pm0.01$&$0.42\pm0.02$&$-1.58\pm0.15$& 2&$-1.87\pm0.02$& 4& 2&  2400\\
J1330+0340&2.8219&2.3215&$21.40^{+0.10}_{-0.10}$&  97&$0.55\pm0.02$&$0.50\pm0.04$&$-1.13\pm0.15$& 2&$-1.45\pm0.06$& 5& 2&  1200\\
J1337-0246&3.0633&2.6871&$20.60^{+0.10}_{-0.10}$&   6&$NA^d$&$0.06\pm0.02$&$-2.80\pm0.13$& 1&$-2.74\pm0.10$& 1& 2&  2400\\
J1339+0548&2.9797&2.5851&$20.60^{+0.10}_{-0.10}$& 175&$0.70\pm0.01$&$0.95\pm0.01$&$-0.91\pm0.11$& 1&$-0.99\pm0.03$& 4& 3&  3600$^{7}$\\
J1340+1106&2.9140&2.7958&$20.85^{+0.10}_{-0.10}$&  40&$0.22\pm0.01$&$0.09\pm0.01$&$-1.70\pm0.10$& 4&$-2.02\pm0.01$& 1& 3& 10800$^{12}$\\
J1344-0323&3.2644&3.1900&$20.95^{+0.10}_{-0.10}$& 285&$1.28\pm0.03$&$0.38\pm0.03$&$-0.61\pm0.19$&15&$-0.91\pm0.10$&11& 1&  2200\\
J1353-0310&2.9745&2.5600&$20.55^{+0.10}_{-0.10}$& 220&$0.96\pm0.01$&$0.79\pm0.01$&$-0.89\pm0.12$& 1&$-1.31\pm0.02$& 1& 3&  5400$^{7}$\\
J1358+0349&2.8888&2.8516&$20.40^{+0.10}_{-0.10}$&   5&$0.04\pm0.01$&$0.10\pm0.01$&$-2.72\pm0.12$& 1&$-2.63\pm0.10$&11& 2&  2400\\
J1402+0117&2.9469&2.4295&$20.30^{+0.10}_{-0.10}$&  64&$0.20\pm0.01$&$0.22\pm0.01$&$-0.89\pm0.15$& 1&$-0.85\pm0.12$& 4& 2&  2400\\
J1450-0117&3.4663&3.1901&$21.20^{+0.10}_{-0.10}$&  65&$0.92\pm0.03$&$1.04\pm0.03$&$-0.76\pm0.15$&13&$-1.23\pm0.05$& 4& 1&  3000\\
J1453+0023&2.5301&2.4440&$20.35^{+0.10}_{-0.10}$&  65&$0.14\pm0.02$&$0.24\pm0.02$&$-1.95\pm0.11$& 1&$-2.46\pm0.05$& 1& 1&  3600\\
J1550+0537&3.1318&2.4159&$20.65^{+0.10}_{-0.10}$&  85&$0.13\pm0.01$&$0.73\pm0.02$&$-2.41\pm0.11$& 1&$-2.43\pm0.05$& 1& 1&  3000\\
J2036-0553&2.5426&2.2804&$21.20^{+0.15}_{-0.15}$&  71&$0.31\pm0.01$&$0.09\pm0.01$&$-1.72\pm0.16$& 1&$-1.71\pm0.10$& 1& 4& 10800\\
J2049-0554&3.1981&2.6828&$20.30^{+0.10}_{-0.10}$& 105&$0.09\pm0.01$&$0.02\pm0.01$&$-2.24\pm0.11$& 1&$-2.54\pm0.13$& 1& 1&  4500\\
J2122-0014&4.0721&3.2064&$20.30^{+0.10}_{-0.10}$& 125&$0.57\pm0.02$&$0.63\pm0.02$&$-0.70\pm0.17$&13&$-1.53\pm0.04$& 1& 1&  3000\\
J2141+1119&2.5091&2.4263&$20.30^{+0.10}_{-0.10}$&  85&$0.12\pm0.01$&$0.23\pm0.01$&$-1.17\pm0.20$&13&$-1.97\pm0.04$& 1& 1&  8100\\
J2154+1102&3.1947&2.4831&$20.70^{+0.20}_{-0.20}$& 205&$0.96\pm0.04$&$0.93\pm0.04$&$-0.90\pm0.15$&13&$-1.03\pm0.10$&11& 1&  2700\\
J2222-0946&2.9263&2.3544&$20.65^{+0.10}_{-0.10}$& 179&$1.23\pm0.01$&$1.51\pm0.01$&$-0.56\pm0.10$& 4&$-0.91\pm0.06$& 1& 4&  3600\\
J2222-0946&2.9263&2.3542&$20.65^{+0.10}_{-0.10}$& 245&$1.21\pm0.01$&$1.47\pm0.01$&$-0.51\pm0.10$& 1&$-0.64\pm0.02$& 1& 1&  3600\\
J2238+0016&3.4674&3.3654&$20.55^{+0.10}_{-0.10}$&  25&$0.12\pm0.01$&$0.02\pm0.01$&$-2.39\pm0.10$& 1&$-1.26\pm0.10$& 4& 1&  3600\\
J2238-0921&3.2594&2.8691&$20.65^{+0.10}_{-0.10}$& 105&$0.53\pm0.01$&$0.27\pm0.01$&$-1.30\pm0.17$& 1&$-1.14\pm0.06$& 4& 1&  2400\\
J2241+1225&2.6307&2.4175&$21.10^{+0.10}_{-0.10}$&  25&$0.39\pm0.01$&$0.16\pm0.01$&$-1.54\pm0.13$& 1&$-2.22\pm0.04$& 1& 1&  3600\\
J2241+1352&4.4480&4.2833&$21.05^{+0.10}_{-0.10}$&  65&$0.49\pm0.02$&$0.22\pm0.02$&$-1.27\pm0.18$&14&$-1.57\pm0.09$& 4& 1&  2700\\
J2315+1456&3.3492&3.2730&$20.30^{+0.10}_{-0.10}$&  85&$0.24\pm0.01$&$0.57\pm0.02$&$-1.76\pm0.11$& 1&$-2.05\pm0.03$& 6& 1&  1800\\
J2334-0908&3.3169&3.0572&$20.40^{+0.10}_{-0.10}$& 212&$0.54\pm0.01$&$0.57\pm0.01$&$-0.98\pm0.10$& 1&$-1.50\pm0.00$& 1& 3& 42720$^{13,11}$\\
J2343+1410&2.9130&2.6768&$20.50^{+0.15}_{-0.15}$&  38&$0.20\pm0.01$&$0.17\pm0.01$&$-1.98\pm0.17$&14&$-2.28\pm0.05$& 6& 4&  3600\\
J2348-1041&3.1724&2.9979&$20.55^{+0.15}_{-0.15}$& 195&$NA^d$&$NA^d$&$-1.80\pm0.15$& 1&$-1.40\pm0.14$& 4& 4&  1800\\
J2350-0052&3.0254&2.6147&$21.25^{+0.10}_{-0.10}$&  93&$0.38\pm0.01$&$0.12\pm0.01$&$-1.92\pm0.10$& 1&$-2.12\pm0.02$& 4& 3& 85500$^{14,15}$\\
J2350-0052&3.0254&2.4269&$20.45^{+0.10}_{-0.10}$& 271&$1.39\pm0.01$&$1.06\pm0.01$&$-0.62\pm0.10$& 1&$-1.08\pm0.00$& 1& 3& 85500$^{14,15}$\\
\hline\hline
\end{tabular}
}

%\label{tab:sample}
\begin{spacing}{0.7}
{\scriptsize { \bf $^a$} Flag describing the metallicity measurement: (1) Si measurement; (2) Zn measurement; (4) = S measurement, (13) mix of limits; (14) Fe measurement + 0.3; (15) Fe Limit + 0.3  } \\
{\scriptsize { \bf $^b$} Flag describing the Fe Measurement: (1) Fe abundance; (2) Fe lower limit; (3) Fe upper limit; (4) Ni abundance offset by -0.1; (5) Cr abundance offset by -0.2; (6) Al abundance; (11) Fe limits from a pair of transitions; (13) Limit from Fe+Ni  } \\
{\scriptsize { \bf $^c$} Instrument Used: 1 = MagE (FWHM$\sim$71 km s$^{-1}$), 2 = XShooter (FWHM$\sim$59 km s$^{-1}$), 3 = UVES (FWHM$\sim$8 km s$^{-1}$), 4 = HIRES (FWHM$\sim$8 km s$^{-1}$) } \\
{\scriptsize {\bf $^d$} NA indicates either no spectral coverage or severe blending that precluded an equivalent width measurement} \\
{\scriptsize {\bf Table Notes.} 
--VLT Program ID Number: 
1: 069.A-0613;
2: 071.A-0114;
3: 073.A-0653;
4: 074.A-0201;
5: 080.A-0014;
6: 080.A-0482;
7: 081.A-0334;
8: 080.A-0482;
9: 076.A-0376;
10: 067.A-0146;
11: 073.B-0787;
12: 067.A-0078;
13: 068.A-0600;
14: 072.A-0346;
15: 079.A-0404
} 
\end{spacing}
\end{table*}

The MagE data were reduced using an IDL reduction package kindly provided by George Becker \footnote{Pipeline available at: \\ ftp://ftp.ociw.edu/pub/gdb/mage\_reduce/mage\_reduce.tar.gz}.  High-resolution spectra from VLT/UVES were reduced using the ESO Common Pipeline Language suite following standard recipes. The extracted spectra from all echelle orders of all exposures were combined using {\sc uves\_popler}\footnote{{\sc uves\_popler} is maintained by MTM at http://astronomy.swin.edu.au/$\sim$mmurphy/UVES\_popler}.  Keck/HIRES spectra were reduced using the XIDL\footnote{http://www.ucolick.org/$\sim$xavier/HIRedux/index.html} reduction package.  X-Shooter data were reduced with the ESO X-Shooter pipeline release 1.2.2.  Continuum fitting of the reduced quasar spectrum was done using the XIDL command x\_continuum, which allows for an interactive spline fit through data points. Because of the difficulty of determining the true continuum blue-ward of the Lyman alpha emission peak -- in the Lyman alpha forest -- we discuss the implications of continuum fit errors in detail in Paper II ~\citep{jorgenson13b} and note that for the metal transitions analyzed here, this is not such a problem.  The atomic data of the transitions discussed in this paper (e.g. laboratory wavelengths, oscillator strengths etc.) were taken primarily from Morton 2003 (Table 2) and the meteoritic solar abundances in Table 1 of ~\cite{asplund09}.  All spectra used in this study are available for download at http://www.dlaabsorbers.info.

\subsection{New \dla\ and Super Lyman Limit System (SLLS) Discoveries}~\label{sec:discoveries}
In the course of this survey we discovered three new \dlas , and three new SLLS, bringing our total \dla\ sample to 109 \dlas , of which 99 have spectra.  While we incorporate the new \dlas\ into the survey, the SLLSs were not included in any further analysis. Details for these new discoveries are as follows: 

$\bullet$ J0011$+$1446 contains a newly discovered \dla\ at z$_{abs}$ = 3.4522 with log \nhi\ = 21.4 $\cm{-2}$ not reported in the SDSS DR5 \dla\ sample ~\citep{pro05} or SDSS DR7 \dla\ sample ~\citep{noterdaeme09}.  The original target in this line of sight is a metal-poor \dla\ ([M/H]=$-$2.5) at z$_{abs}$ = 3.6175. 

$\bullet$ J1019$+$0825 contains a SLLS at z$_{abs}$ = 2.9653 with log \nhi\ = 20.15 $\cm{-2}$ and a SLLS at z$_{abs}$ = 2.4373 with  log \nhi\ = 19.80 $\cm{-2}$.

$\bullet$ J1111$+$1332 contains a \dla\ at z$_{abs}$ = 2.271 with log \nhi\ = 20.50 $\cm{-2}$ and a relatively low metallicity of [M/H] = $-$2.49 that was not reported in the DR5 sample, but was included in the DR7 sample. ~\cite{noterdaeme09} report this \dla\ at z$_{abs}$=2.273, with log \nhi\ = 20.55 $\cm{-2}$.  This line of sight also contains the previously detected SDSS DR5 \dla\ at z$_{abs}$ =  2.3822 (which is not included in the DR7 sample).
 
$\bullet$ J1220$+$0921 contains a newly discovered proximate \dla\ with log \nhi\ = 20.80 $\cm{-2}$ at z$_{abs}$ = 4.1223 not reported in DR5 or DR7.  Because this system is likely associated with the quasar (z$_{em}$ = 4.11027) we do not include it in our sample analysis.  
      
$\bullet$ J1233$+$1100 contains a newly discovered \dla\ at z$_{abs}$ = 2.8206 with log \nhi\ = 20.35 $\cm{-2}$ and [M/H] = $-$1.29 not reported in DR5 or DR7.  The original DR5 \dla\ in this line of sight is at z$_{abs}$ =  2.7924.

$\bullet$ J1246$+$1113 contains a SLLS at z$_{abs}$ = 2.6368  with log \nhi\ = 20.20 $\cm{-2}$ and [M/H] = $-$1.32, not included in the survey because its \nhi\ falls just below the \dla\ cut.  This system was reported in the DR7 sample at  z$_{abs}$ = 2.636 with log \nhi\ = 20.19$\cm{-2}$.  The original DR5 target \dla\ at z$_{abs}$ = 3.0981 is included in the sample (however, was not included in DR7 sample).

\begin{table}
\centering
\caption{ Missing Target DLAs.}
{\scriptsize \begin{tabular}{ccccc}
\hline\hline
Quasar & $z_{abs}$ & log N(HI) & $z_{em}$ & $i$-band Magnitude \\
     &            &        [cm$^{-2}$]          &  &  \\
\hline\hline
J1301$+$1246 & 3.0251 & 20.95 & 4.103590 & 18.774 \\
J1305$+$0521 & 3.6415 & 20.30 & 4.086680 & 18.702 \\
J1305$+$0521 & 3.6790 & 21.10 & 4.086680 & 18.702 \\
J1325$+$1255 & 3.5497 & 20.40 & 4.140640 & 18.842 \\
J1341$-$0303 & 2.7556 & 20.40 & 3.222270 & 18.962 \\
J1341$+$0141 & 3.6330 & 20.65 & 4.670000 & 18.899 \\
J1347$+$0213 & 3.2070 & 20.50 & 3.325270 & 18.985 \\
J1353$-$0250 & 2.3624 & 20.30 & 2.411540 & 18.596 \\
J1402$+$0146 & 3.2773 & 20.95 & 4.160920 & 18.263 \\
J1452$+$0154 & 3.2529 & 21.45 & 3.908270 & 18.685 \\
\hline\hline 
\end{tabular}
}
\label{tab:missing}
\end{table}

%%F3
\begin{figure*}
\includegraphics[width=1.75\columnwidth]{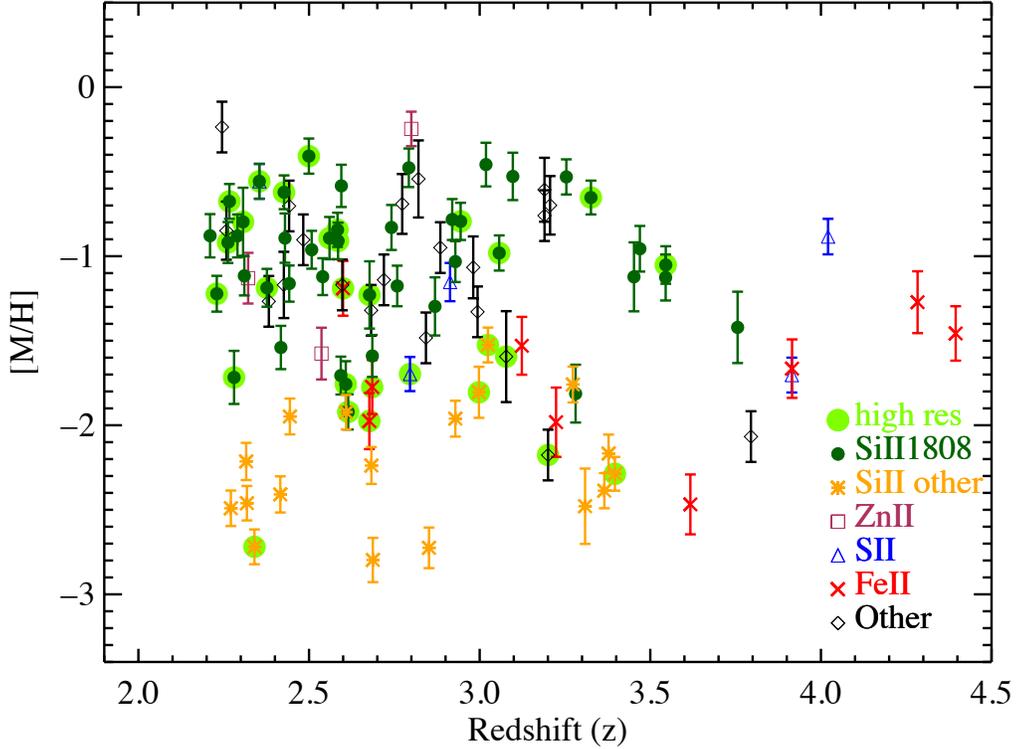}\\
\caption{
[M/H] versus redshift for the Magellan \dla\ sample with the ion used to determine metallicity indicated.  These metallicity measurements have been made after applying the saturation criterion of F$_{min}$/F$_q <$ 0.65 but include no additional corrections based on flux. High-resolution data (either HIRES or UVES) are indicated by the larger chartreuse circle and have no corrections for saturation.  The black diamonds indicated by `Other' are determined by a combination of limits, while orange circles labeled `SiII other' are determined by either Si{\sc \,ii} $\lambda$1304,  Si{\sc \,ii} $\lambda$1526 or an average of the two.
}
\label{fig:mvz}
\end{figure*}

\section{Column density and metallicity measurements}~\label{sec:measure}
     
In this section we describe the measurement of the neutral hydrogen column density, N(HI), the column densities of various metal species expected to dominate the total metal column densities in DLAs, and the metallicity, [M/H]\footnote{We use the standard shorthand notation for metallicity relative to solar, [M/H] = log(M/H) - log(M/H)$_{\sun}$}. We also conduct several tests to explore the effect of limited spectral resolution on our metallicity measurements.
     
\subsection{N(\hi ) and [M/H] Measurements}
We used the x\_fitdla routine contained within the XIDL package to measure the neutral hydrogen column density, \nhi , for each \dla .  This routine allows the user to interactively and simultaneously fit both a Voigt profile to the \dla\ and the surrounding continuum centered on the chosen redshift, with the goal of fitting both the core and the wings of the profile.  Redshifts were determined by the centroid of the strongest, unsaturated, low-ion metal component, or in other words, the low-ion velocity component with the largest optical depth without being saturated.  As pointed out by ~\cite{rafelski12}, this `not completely quantitative' fitting-by-eye method is justified because the errors are dominated by systematic errors attributed to continuum fitting and line-blending.  Following the standard practice, such as that used by ~\cite{pro05} and ~\cite{pro09}, we place conservative error estimates on \nhi\ of a minimum of 0.1 dex.  Figure~\ref{fig:sdssnhi} presents a comparison of the original SDSS determined log \nhi\ values and the Magellan sample, i.e. MagE or higher resolution spectra, determined log \nhi .  The least squares best-fit slope through the data, indicated by the black dashed line, has slope = 1.009, not very different from slope = 1, the solid red line.  

We used the standard apparent optical depth method (AODM; ~\cite{savage91}) to derive the column density of every available metal species in each \dla . Metallicities were typically  determined from the Si II $\lambda$1808  line if available.  In Table~\ref{tab:sample} columns 8 and 9, we report the derived metallicity and a descriptive flag, respectively, for each object.  

When deriving metallicities from medium-resolution (FWHM $\sim$70 \kms ) spectra, one  
possible pitfall can lead to an underestimation of the metallicity due to the potential saturation of unresolved components. In \S~\ref{sec:metals}, we describe the efforts we have made to ameliorate these effects. 

In Figure ~\ref{fig:mvz} we plot the metallicities of the entire sample versus the \dla\ absorption redshift.  Different symbols/colors indicate the ion used to determine [M/H].  Targets for which high-resolution spectra were available are indicated by larger charteuse circles.  It is immediately seen by eye that there is not a strong apparent evolution with redshift.  We provide a detailed discussion of potential redshift evolution and comparison with other \dla\ surveys in \S~\ref{sec:redshiftevol}.

 \subsection{The Influence of Medium-Resolution Spectra in Determining [M/H]:  Applying Flux-based Saturation Corrections}~\label{sec:metals}

The potential saturation %`washing-out' 
of unresolved spectral features has long been a known issue confronting spectral observations of absorption lines. In the context of \dlas\ and the determination of \dla\ gas metallicity, this issue was first discussed by ~\cite{pro03esi}, and later by ~\cite{penprase10}, who demonstrated that measuring metallicities from medium-resolution spectra, particularly in the high equivalent width regime, can lead to an underestimation of the true metallicity because of a failure to account for unresolved and potentially saturated spectral components. ~\cite{pro03esi} found that in a comparison of Keck ESI (R $\sim$ 7,000) and Keck HIRES (R $\sim$ 50,000) spectra, the observed differences in the derived column densities were significant when the equivalent width was $>$1\AA .  \cite{penprase10}, who conducted a survey for metal-poor \dlas , analyzed simulations of a single OI $\lambda$1302 line and concluded that in the equivalent width range of $W$ = 0.050 \AA $-$ 0.130\AA , a saturation correction should be applied, whereas at $W >$ 0.130\AA , the line must be considered saturated.  

 %%F4
 \begin{figure}
\includegraphics[width=0.99\columnwidth]{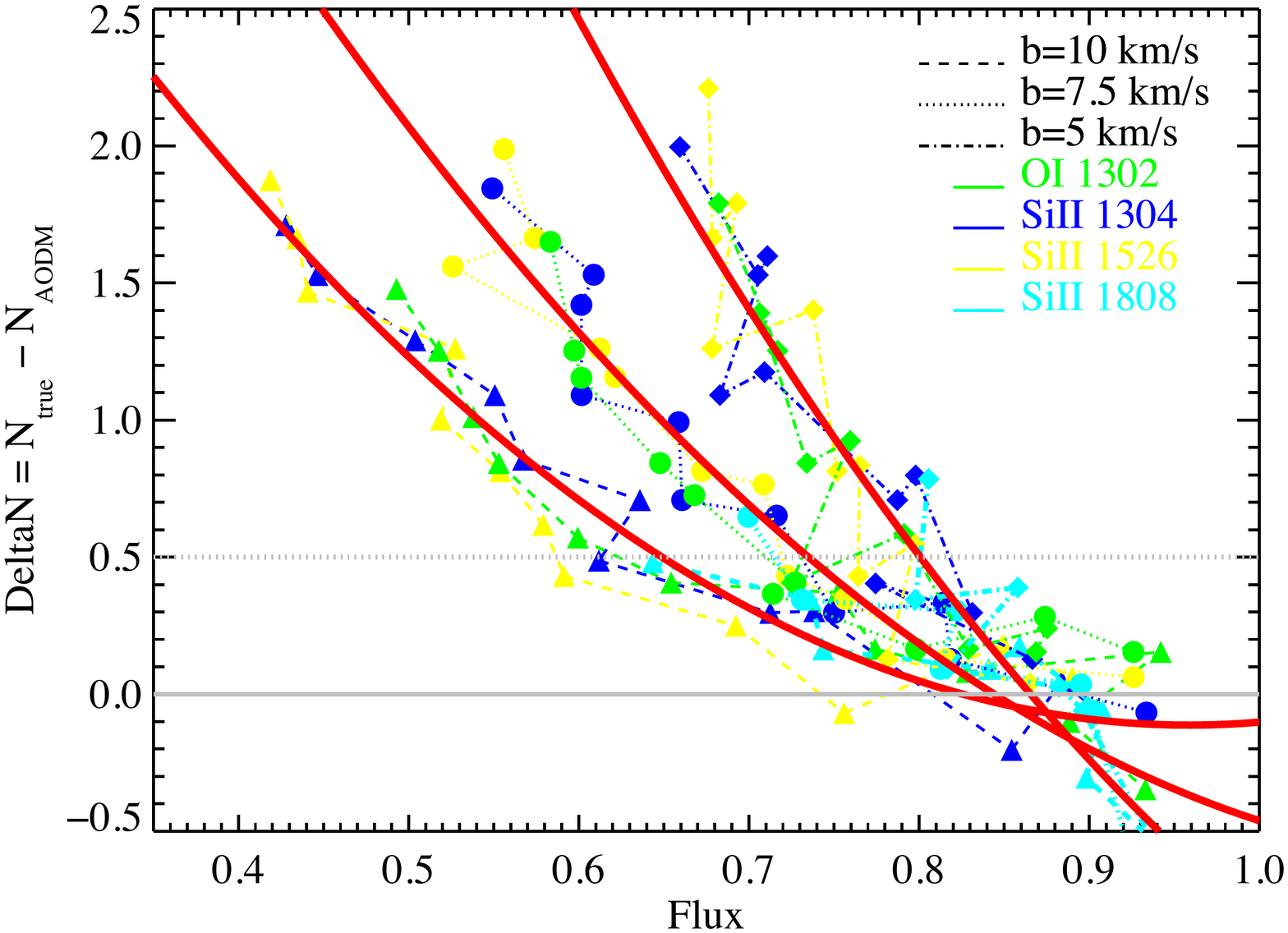}\\
\caption{
Results of simulations to estimate the effects of medium-resolution MagE spectra on metallicity determination.  Using simulated spectra that match those of the MagE data we inserted and measured lines of OI and SiII with a range of column densities and Dopper b parameters.  The above figure plots $\Delta$N, the difference between the true/input column density (N$_{true}$) and the output column density (N$_{AODM}$) measured by the AODM technique versus the minimum flux of the absorption line.  The different colors represent the different ion transitions used and the symbols and linestyles connecting the points indicates the Doppler parameter assumed.  Overplotted as a solid red line is a best-fit polynomial curve to the data for b = 10, 7.5 and 5 km s$^{-1}$, from left to right.  This fit can be used to make corrections to the derived column density of the ion used to determine [M/H] based upon the minimum flux within the profile.    
}
\label{fig:deltan}
\end{figure}

Given the MagE resolution of FWHM$\sim$ 71 \kms\ (R $\sim$ 4100), we have attempted to account for the possibility of unresolved saturation in our determination of metallicities from the MagE spectra.  In the following sections we report on several tests we performed in order to determine the best possible corrections, if any, to make.

%%F5
\begin{figure}
\includegraphics[width=0.99\columnwidth]{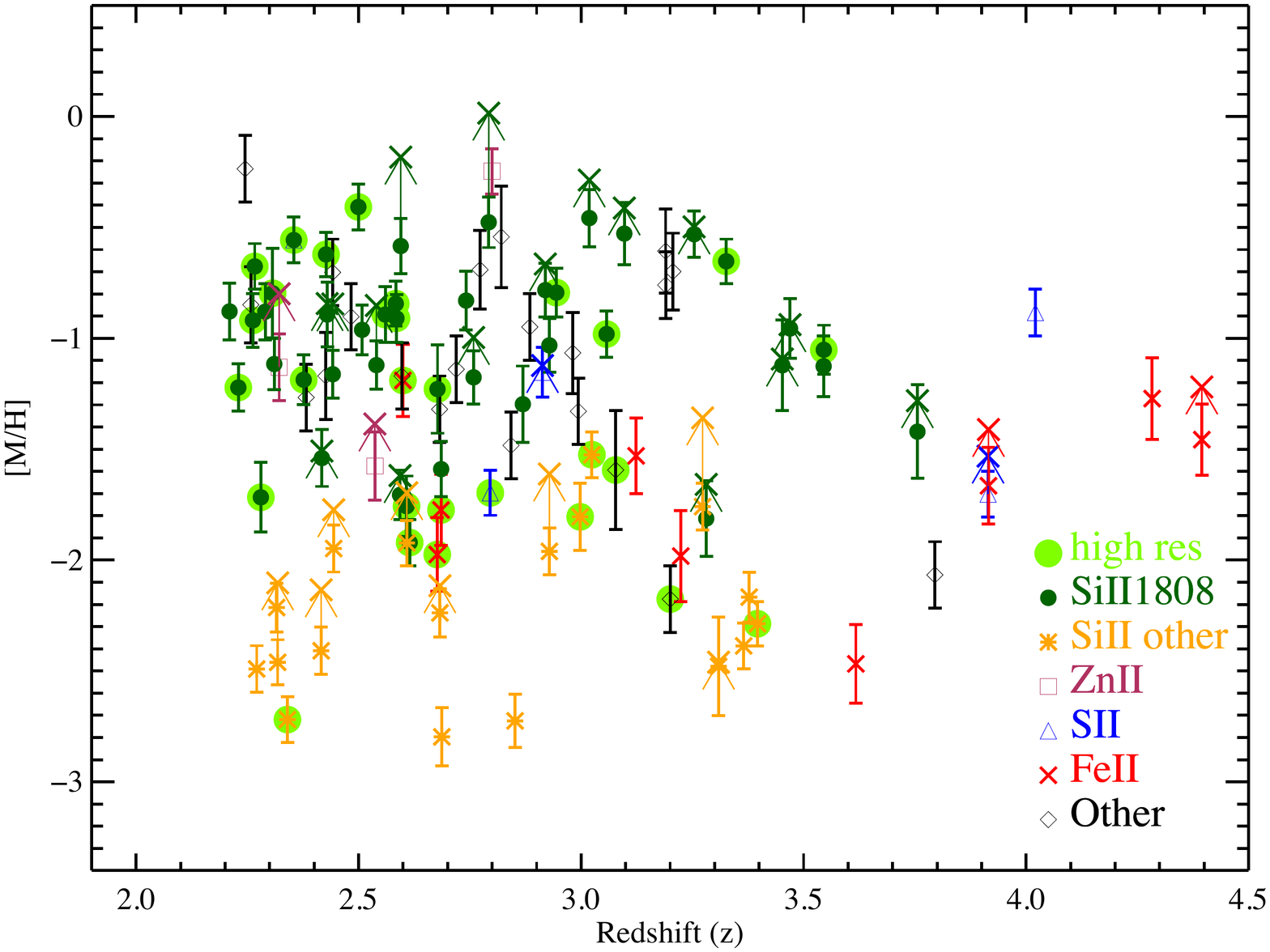}\\
\caption{
[M/H] versus redshift for the Magellan \dla\ sample with the ion used to determine metallicity indicated. Flux corrections assuming b = 10 \kms\ have been applied.  The `X' indicates the new flux-corrected metallicity and the arrow connects the original to new point. 
}
\label{fig:mvz_corrected}
\end{figure}

\subsubsection{Corrections Test 1: Simulated Data}
We note that saturation determined purely from an equivalent width measurement can be misleading in the case of wide, complex velocity profile systems.  Rather, we define, as done by ~\cite{herbert06}, a normalized flux level below which an absorption line is considered saturated.  ~\cite{herbert06} found that normalized intensities F$_{min}$/F$_q <$ 0.4, where F$_{min}$ is the minimum absorbed flux and F$_q$ is the unabsorbed quasar flux, are typically saturated in Keck/ESI data.  

To determine this level for the Magellan/MagE spectra we simulated spectra to match the typical observed MagE spectra, i.e. resolution $\sim$ 71 \kms\  and S/N$\sim$30.  We analyzed single velocity component absorption lines of OI\ $\lambda$1302, SiII\ $\lambda$1304, SiII\ $\lambda$1526, and SiII\ $\lambda$1808 with a range of column density values and Doppler parameters, to determine the flux level above which the AODM returned a column density value, N$_{AODM}$, similar to the true (i.e. input) value, N$_{true}$.  The results of these simulations are shown in Figure~\ref{fig:deltan}, where we plot $\Delta$N, the difference in N$_{true}$ and N$_{AODM}$, versus the minimum flux of the absorption line, for three different Doppler parameters, b = 10, 7.5 and 5 km s$^{-1}$.  While there is a large variation depending on the line and column density/Doppler parameter chosen, we note that for b = 10 km s$^{-1}$, any absorption line with a normalized flux of greater than F$_{min}$/F$_q >$ 0.65 was prone to deviations from N$_{true}$ of 0.5 dex or less. We take b = 10 \kms\  to be a representative Doppler parameter because, while it might be slightly larger than a typical individual velocity component, most low-ion profiles contain blends of several velocity components qualitatively similar to larger Doppler parameter profiles. Therefore, we have implemented an automatic saturation criterion that flags any line where F$_{min}$/F$_q <$ 0.65 as saturated and not to be used to determine [M/H].

In addition, we used the results of the simulations to attempt to apply flux-based saturation corrections to the lines used to determine [M/H] in the MagE spectra. Shown as the red solid lines in Figure~\ref{fig:deltan} is the best-fit polynomial curve to the data for b = 10, 7.5 and 5 \kms (from left to right).  We then applied this fit to make corrections to the derived column density of the ion used to determine [M/H] based upon the minimum flux within the profile.  It is apparent from Figure~\ref{fig:deltan} that as the Doppler parameter becomes smaller, the necessary column density corrections can become rather significant.  While we take note that this is an issue to be aware of, using the  justification stated above, we apply the corrections based on the best-fit line for Doppler parameter b = 10 km s$^{-1}$.  In Figure~\ref{fig:mvz_corrected} we show the resulting changes to the metallicity versus redshift after applying this flux-based correction for b = 10 \kms\ to the MagE and X-Shooter data.  We note that applying these flux-based metallicity corrections does not make a large change in the final result -- the mean correction of the lines that required a correction is only 0.18 dex. Therefore, given the uncertainties and variations in Doppler parameters, along with the fact that this exercise indicates these will not make a large difference, we decided not to apply this flux based correction to the sample.

%%F6
\begin{figure}
\includegraphics[width=0.99\columnwidth]{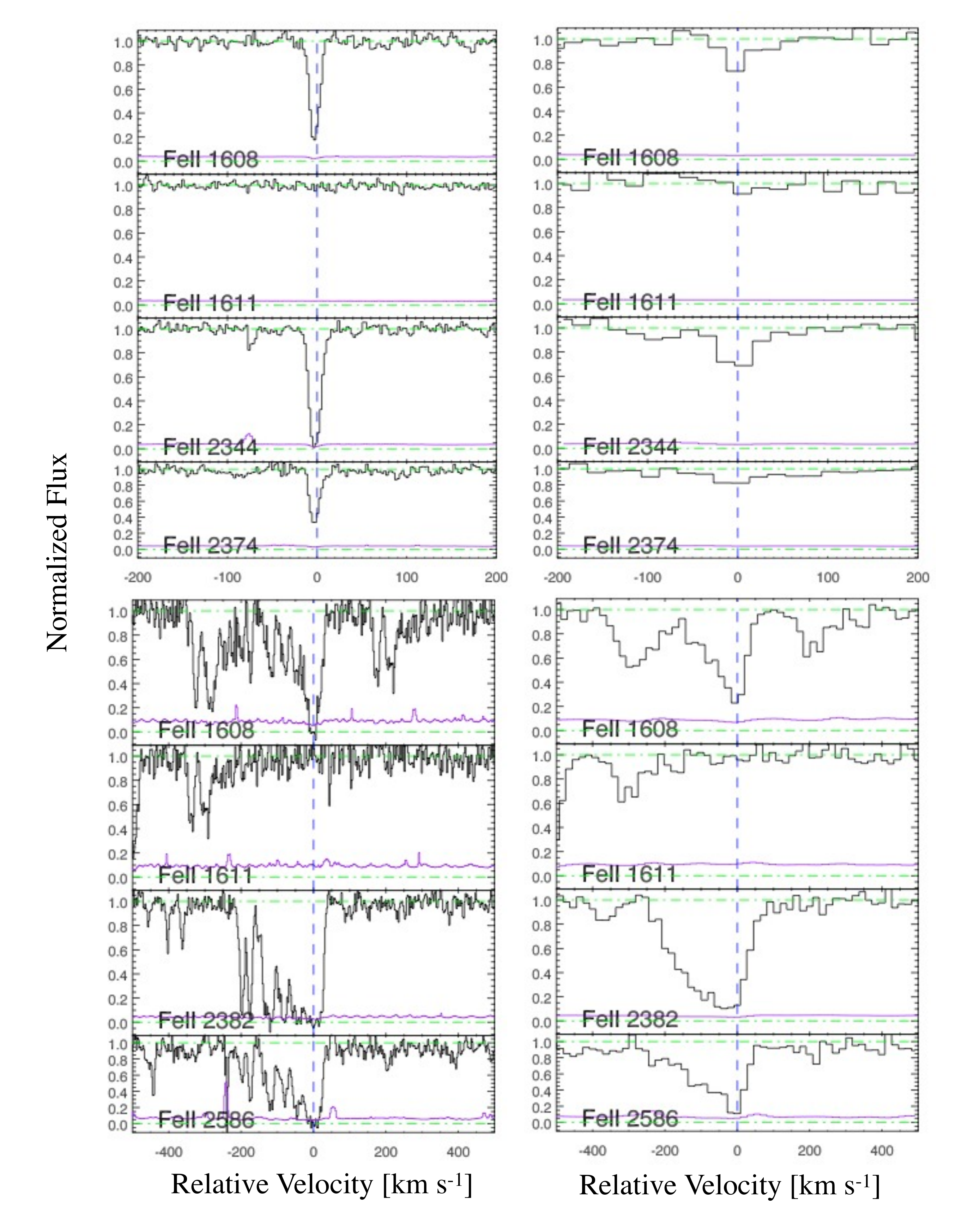}\\
\caption{
 Example Fe{\sc \,ii} velocity profiles used to create Figure~\ref{fig:minfluxsmooth}.  The top two panels show a selection of Fe{\sc \,ii} velocity profiles from \dla\ 1155$+$0530.  On the left is the original UVES spectrum (FWHM $\sim$8 km s$^{-1}$) and on the right is the same spectrum smoothed to MagE resolution (FWHM$\sim$70 km s$^{-1}$) at a SNR = 20.  The bottom two panels show a selection of Fe{\sc \,ii} velocity profiles for \dla\ 0338$-$0005, with the original UVES spectrum on the left and the smoothed spectrum on the right.
}
\label{fig:fig5example}
\end{figure}

\subsubsection{Corrections Test 2: Smoothed high-resolution Data}
\begin{figure*}
\includegraphics[width=1.75\columnwidth]{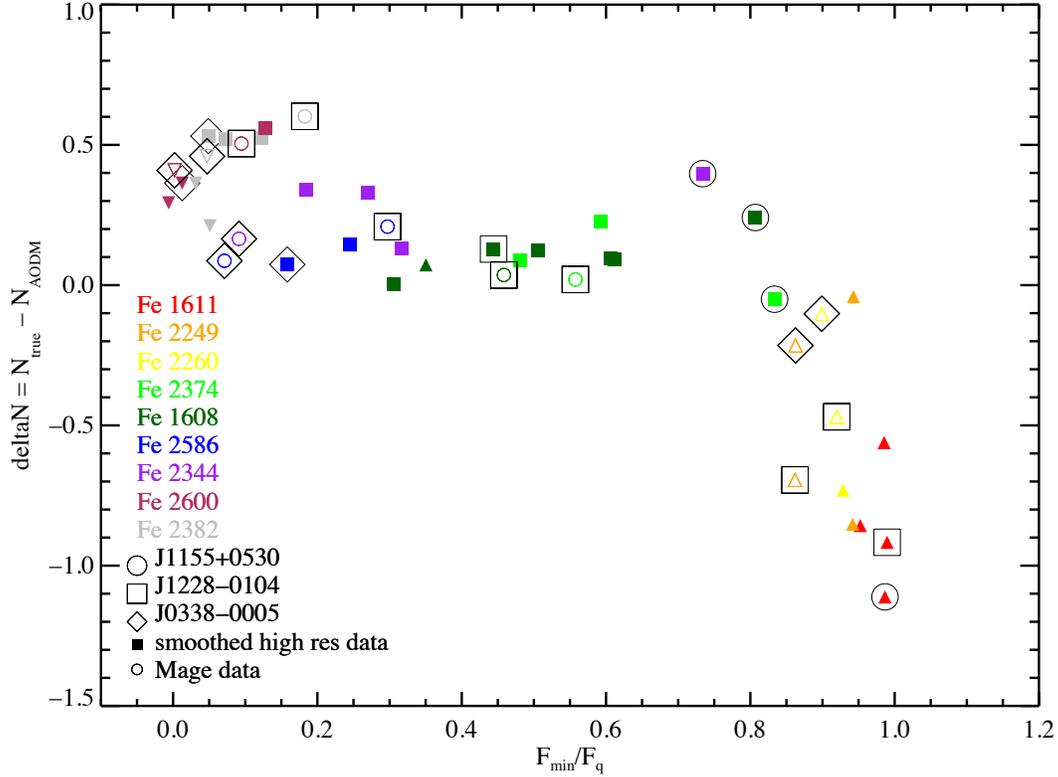}\\
\caption{The results of Corrections Test 2 plotted as $\Delta$N versus F$_{min}$/F$_q$.  Colors indicate different Fe{\sc \,ii} lines in rainbow order of increasing oscillator strength.  Triangles indicate that the AODM measurement was an upper limit (upward pointing) or lower limit (downward pointing).  Open circles represent data from \dlas\ that also had MagE spectra (so these measurements are based on the MagE spectrum).  Filled squares are the data derived from the smoothed spectrum.  Larger open symbols denote the particular object, i.e. dark green square and dark green circle located at F$_{min}$/F$_q$ $\sim$ 0.45 are both inside a larger black square, indicating they are the Fe{\sc \,ii} $\lambda$1608 line from the smoothed UVES J1228$-$0104 spectrum and the original MagE J1228$-$0104 spectrum respectively.  
}
\label{fig:minfluxsmooth}
\end{figure*}

In reality, the low-ion velocity structure of \dlas\ usually consists of many velocity components with various degrees of blending, rather than a simple, single velocity component.  In light of this fact, we performed a similar analysis as above, however, this time instead of using simulated data at MagE resolution, we used high-resolution data from our sample and smoothed it to the resolution of the MagE spectra, adding noise such that the SNR of the smoothed spectrum matched that of our typical MagE spectra, SNR$\sim$20 pixel$^{-1}$.  We choose the eight high-resolution spectra that contained coverage of the large range of Fe{\sc \,ii} transitions from $\lambda$1608\AA\ $-$ $\lambda$2600\AA .  In this way we were able to 1) have a good measure of the true N(Fe{\sc \,ii}), N$_{true}$, from the unsaturated high-resolution Fe{\sc \,ii} transitions, and 2) explore the effects of measuring N$_{AODM}$ from a wide range of oscillator strength transitions in the context of realistic velocity profiles smoothed to the resolution of MagE.  Figure~\ref{fig:fig5example} contains example Fe{\sc \,ii} velocity profiles for two of these \dlas , \dla\ 1155$+$0530 and \dla\ 0338$-$0005. The left column is the original high resolution UVES spectrum, while the right column is the same spectrum smoothed to the MagE resolution with noise added such that SNR$\sim$20 pixel$^{-1}$.  In Figure~\ref{fig:minfluxsmooth} we show that, perhaps surprisingly, the effect of underestimation of the true column density may not be so severe in reality.  

%%F7

\begin{table}
\centering
\caption{MagE versus high resolution.}
{\scriptsize \begin{tabular}{ccccc}
\hline\hline
DLA & feature & MagE & high res. & $\Delta$$_{MagE - high res}$  \\
\hline\hline
\dla\ 0035$-$0918$^a$ & [M/H] & $-$2.61$\pm$0.12 &$-$2.72 $\pm$0.10 & 0.11 \\
& \delvninty\ (\kms ) & 25 & 22 & 3 \\
& W$_{1526}$ (\AA )  & 0.06 $\pm$ 0.25 & 0.04 $\pm$ 0.05  &  0.02\\
\dla\ 0338$-$0005$^b$ & [M/H] &$-$1.34$\pm$0.13 & $-$1.22$\pm$0.11 & 0.12 \\
& \delvninty\ (\kms ) & 165 & 227 & $-$62 \\
& W$_{1526}$ (\AA )   &1.32 $\pm$0.44 &1.12$\pm$0.14 & 0.20\\
\dla\ 1057$-$0629$^b$ & [M/H] &$-$0.36$\pm$0.11 & $-$0.41$\pm$0.10 & 0.05 \\
& \delvninty\ (\kms ) & 405 & 253 & 152 \\
& W$_{1526}$ (\AA )  & 1.65 $\pm$0.55&1.57$\pm$0.18 & 0.08 \\
\dla\ 1228$-$0104$^b$  & [M/H] &$>$ $-$0.90 & $-$0.92$\pm$0.12 & 0.02\\
& \delvninty\ (\kms )& 125 & 98 & 27 \\
& W$_{1526}$ (\AA ) & 0.64 $\pm$0.43 &0.44 $\pm$0.13  & 0.20\\
\dla\ 2222$-$0946$^a$ & [M/H] & $-$0.52$\pm$0.10 &$-$0.56 $\pm$0.10 & 0.04\\
& \delvninty\ (\kms ) & 245 & 179 & 66 \\
& W$_{1526}$ (\AA )  & 1.23 $\pm$ 0.47 & 1.22 $\pm$ 0.01  &  0.01\\
\hline\hline 
\end{tabular}
}
\label{tab:mage_hires_comp}
\begin{spacing}{0.7}
{\scriptsize {\bf $^a$} Keck/HIRES data } \\
{\scriptsize {\bf $^b$} VLT/UVES data  } \\
\end{spacing}
\end{table}

%%F8
\begin{figure}
\includegraphics[width=0.99\columnwidth]{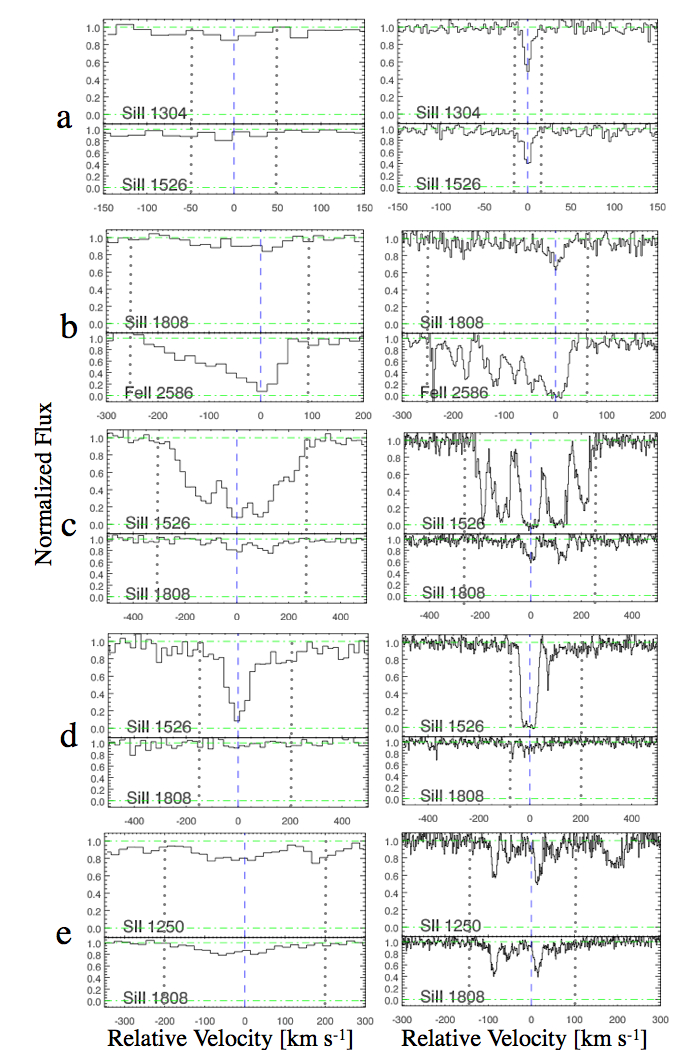}\\
\caption{
Low-ion velocity profiles used to estimate the metallicity, [M/H], in each \dla\ in Table~\ref{tab:mage_hires_comp}. The panels, labelled a$-$e represent \dlas\ 0035$-$0918, 0338$-$0005, 1057$-$0629, 1228$-$0104 and 2222$-$0946, respectively.  Each panel shows the MagE spectrum on the left and the high resolution spectrum on the right.  In all panels, the transition used to estimate the metallicity is shown, as well as another transition for reference.  The grey dotted lines denote the velocity range over which the metallicity was calculated. In panels b, c, and d the Si{\sc \,ii} $\lambda$1808 line was used to determine the metallicity, while in panel a, both Si{\sc \,ii} $\lambda$1304 and Si{\sc \,ii} $\lambda$1526 were used, and in panel e, Si{\sc \,ii} $\lambda$1808 was used in the case of the MagE spectrum, while S{\sc \,ii} $\lambda$1250 was used for the high resolution spectrum.  
}
\label{fig:table3fig}
\end{figure}

In Figure~\ref{fig:minfluxsmooth} we plot $\Delta$N versus F$_{min}$/F$_q$ for eight \dlas , where $\Delta$N = N$_{true}$ - N$_{AODM}$.  N$_{true}$ is measured from the non-saturated Fe{\sc \,ii} transitions in the high-resolution data, while N$_{AODM}$ is measured for each of the Fe{\sc \,ii} transitions in the smoothed spectrum.  Note that the transitions of large unabsorbed flux (right side of plot) are associated with weak transitions for which upper limits on N$_{AODM}$ were returned, making the resultant $\Delta$N negative.  The large black circle denotes the Fe{\sc \,ii} transitions belonging to \dla\ 1155$+$0530, a somewhat unusual case, as it contains a single relatively narrow absorption feature (see Figure~\ref{fig:fig5example}, where the velocity width of 90\% of the optical depth, \delvninty\ = 27 km s$^{-1}$) and therefore probes a specific range in flux.  However, it is seen from these lines that even in this case of a narrow absorption feature, the MagE-resolution AODM does not greatly underestimate the column density.

From this analysis we conclude that over a range of oscillator strengths, for a range of real, low-ion velocity profiles, the underestimation caused by application of the AODM technique to medium-resolution data is generally less than $\sim$0.5 dex, and for some oscillator strength/flux/column density combinations is less than 0.3 dex.  Interestingly, even in the range of flux F$_{min}$/F$_q \sim$0.4$-$0.6 the underestimation is not large.  These results give us confidence that our choice of minimum flux F$_{min}$/F$_q <$ 0.65 as the definition of a saturated transition is a conservative limit that will ensure we are making a minimum amount of metallicity underestimation due to the application of AODM on medium-resolution data.

\subsubsection{Corrections Test 3: Direct Comparison of MagE and High-resolution Data}
 As an additional check on the robustness of the MagE metallicities, we compared the results for the five sample \dlas\ observed with both MagE and a high-resolution spectrograph.  We summarize the results in Table~\ref{tab:mage_hires_comp} and make note of the fact that, by chance, these five \dlas\ happen to span a large range in metallicity, \delvninty\ and equivalent width of the Si{\sc \,ii} $\lambda$1526 ($W_{\lambda1526}$) transition.  In Figure~\ref{fig:table3fig} we plot the velocity profile of the low-ion used to estimate the metallicity, as well as another transition for reference, for each \dla\ in Table~\ref{tab:mage_hires_comp}. It is reassuring to see from this comparison that the derived [M/H] agree quite well -- with a maximum difference of 0.12 dex.     
  
We note that for three \dlas ,  \dla\ 0338$-$0005, \dla\ 1057$-$0629 and \dla\ 2222$-$0946, there is a relatively large discrepancy in the \delvninty\ measurement of the MagE versus high resolution spectrum.  As described in \S~\ref{sec:delv}, the way in which \delvninty\ is measured, by moving pixel-by-pixel across the velocity profile, means that it is dependent upon the resolution of the spectrum, and, as a result, lower resolution spectra tend to overestimate the true \delvninty .  While we have attempted to correct for this effect, as discussed in \S~\ref{sec:delv}, it is clear that perhaps, in the case of these 3 \dlas , the correction is not enough. In addition, we point out that in these cases, the velocity interval over which the \delvninty\ is calculated is different for the different resolution spectra, as seen in Figure~\ref{fig:table3fig}.  If we instead calculate the \delvninty\ in the MagE spectrum using the same velocity interval as that used for the high resolution spectrum, we find that for two \dlas, the discrepancies in \delvninty\ become smaller.  Specifically, \delvninty $^{MagE}$ = 305 km s$^{-1}$ and 205 km s$^{-1}$ for \dla\ 1057$-$0629 and \dla\ 2222$-$0946, respectively.  On the other hand, DLA 0338$-$0005 became slightly more discrepant when integrated over the high resolution velocity profile, with \delvninty\ = 125 km s$^{-1}$.  As a result, we stress that one must use caution when interpreting the measurements of \delvninty\ from lower resolution spectra.  
  
  %%F9
\begin{figure}
\includegraphics[width=0.99\columnwidth]{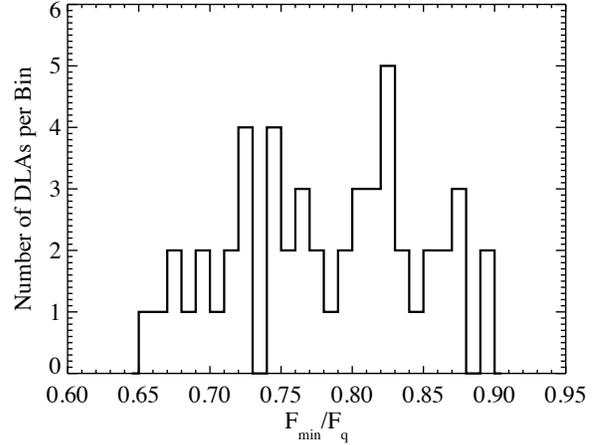}
\caption{Distribution of F$_{min}$/F$_q$ values of the transitions used in determining the metallicities for 51 of the MagE and X$-$Shooter spectra.  The remaining 19 were were determined from a combination of limits and are not shown here.
}
\label{fig:magehist}
\end{figure}

%%F10
\begin{figure*}
\includegraphics[width=1.75\columnwidth]{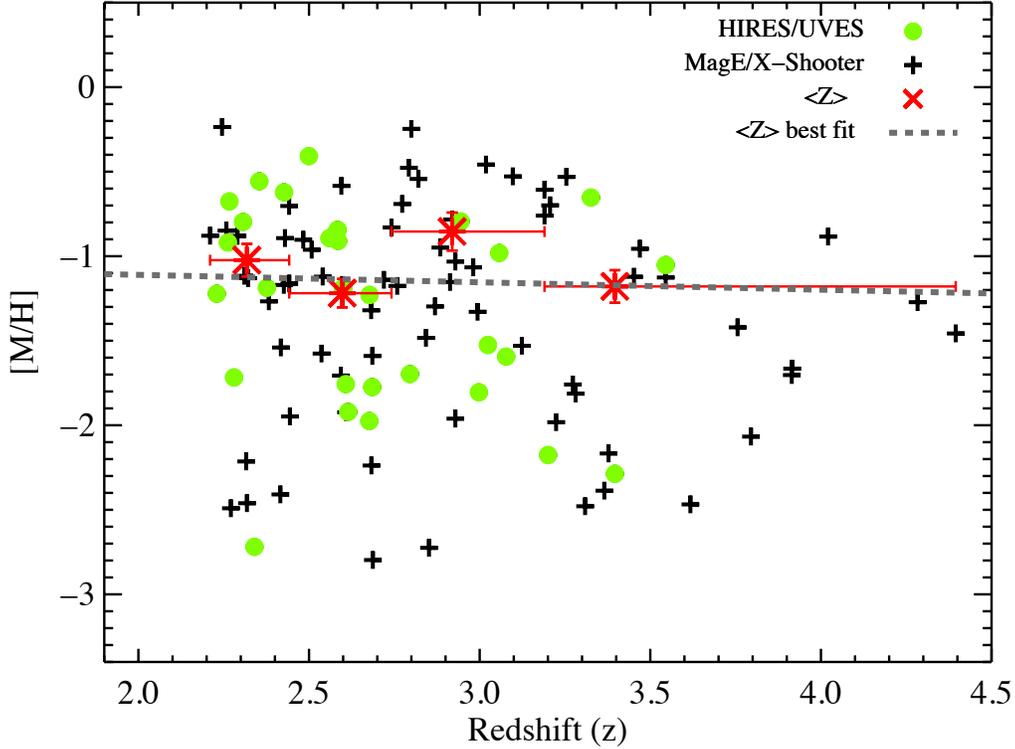}\\
\caption{
[M/H] versus redshift for the Magellan \dla\ sample.  The sample is divided into four redshift bins containing equal numbers of \dlas .  The red points denote the cosmic mean metallicity with 1$\sigma$ confidence interval bootstrap error bars as defined in the text. The grey dashed line represents the linear best-fit to the binned data points, $\langle Z \rangle$ = (\fitcosmob )$z -$(1.06$\pm$0.36).
}
\label{fig:mvz_cosmomean}
\end{figure*}

\subsubsection{Summary of Corrections Tests and Final Adopted Metallicities}
Taking the results of these three corrections tests into consideration, we conclude that with the possible exception of some rare cases, resolution-motivated corrections to the AODM derived column densities from the medium-resolution MagE data will not make a significant difference to the results, provided we apply the minimum flux requirement of F$_{min}$/F$_q <$ 0.65.  Therefore, we report in Table~\ref{tab:sample} the AODM-derived metallicities of the MagE and X-Shooter data without additional flux-based corrections unless otherwise noted.  In Figure~\ref{fig:magehist} we show the distribution of F$_{min}$/F$_q$ values of the transitions used in determining the metallicities for 51 of the MagE and X$-$Shooter spectra. It is clear that the F$_{min}$/F$_q$ values are not clustered at the set threshold (0.65) and therefore, we expect that our results do not strongly depend on the threshold value we use.  An additional 19 systems, labeled in Figure~\ref{fig:mvz} as `Other' and determined from a combination of limits, are not shown.

\section{Cosmic Metallicity Evolution over Redshifts $\lowercase{z}$ = 2.2 $-$ 4.4?}\label{sec:redshiftevol}
 In this section we investigate the evidence for redshift evolution in the \dla\ metallicities of the Magellan sample. 
We compare our sample to that of ~\cite{rafelski12} who found a metallicity evolution of $-$0.22 $\pm$ 0.03 dex per unit redshift over the redshift range $z$ = 0.09 $-$ 5.06.   Rafelski et al. (2012, hereafter R12) report metallicity measurements for 47 new \dlas , 30 with z$_{abs}>$ 4, and incorporate an additional 195 \dlas\ from the literature.  Their literature sample includes a sample created earlier by 
~\cite{pro03met} who -- for the first time detected a statistically significant evolution in the cosmic mean metallicity of $-$0.26 $\pm$ 0.07 dex per unit redshift.  The ~\cite{pro03met} sample consists of 125 \dlas , $\sim$ 75 of which were drawn from the literature, while the other $\sim$ 50 were taken by them with ESI.  

%%F11
\begin{figure*}
\includegraphics[width=1.75\columnwidth]{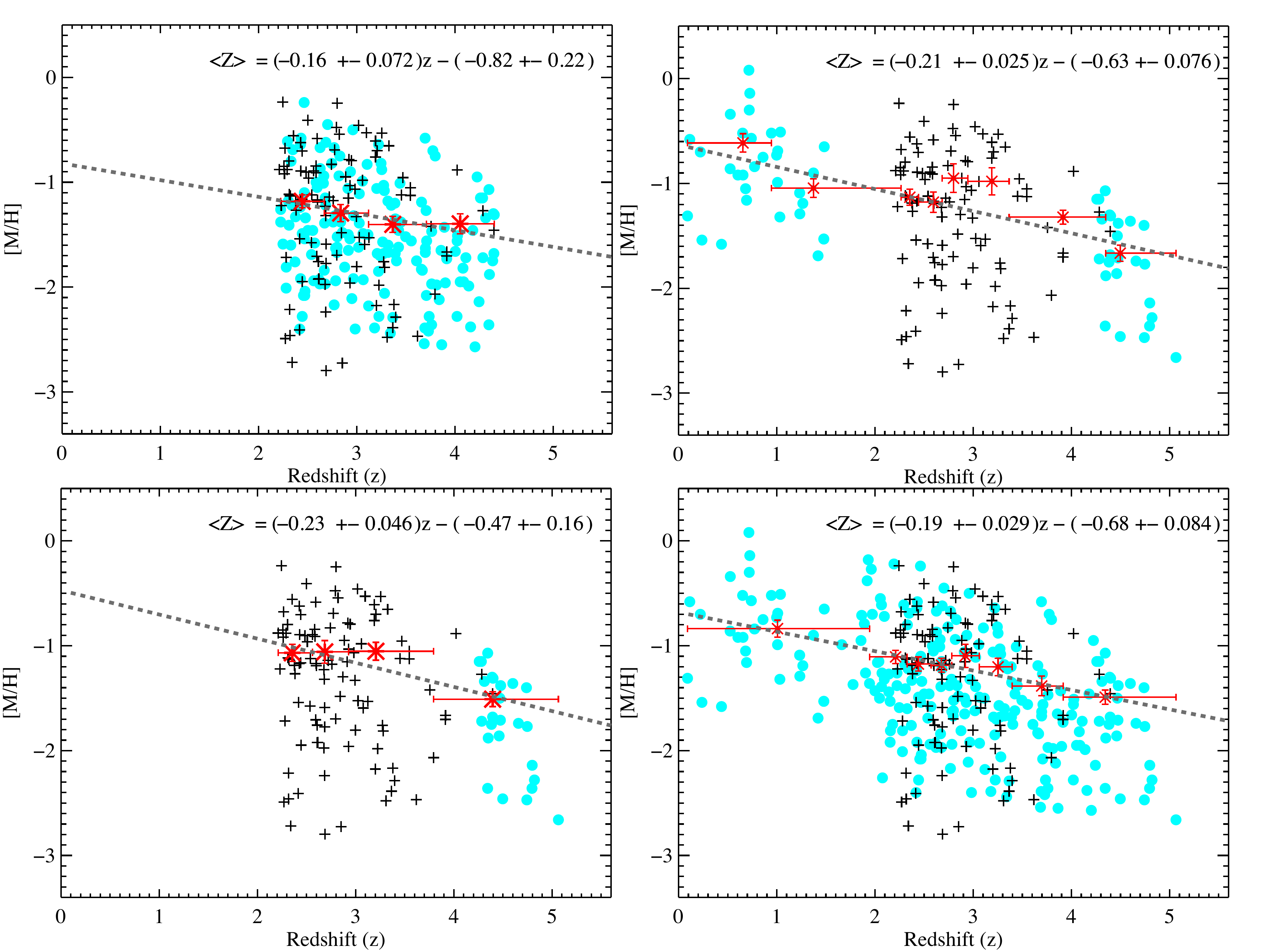}\\
\caption{[M/H] versus redshift for the Magellan sample (black crosses) and various subsets of the \hiz\ sample (cyan circles).  Overplotted in red are the cosmological mean values in either 4 or 8 bins of equal numbers of \dlas\ with 1$\sigma$ error bars determined as described in the text.  The dashed grey lines are linear fits to the $\langle$Z$\rangle$ data points.  {\bf Top left panel:} The \hiz\ sample taken over only the redshift range of the Magellan sample (Magellan sample points shown only for reference).  It is seen that when the \hiz\ sample is constrained to this smaller redshift range the slope is flatter ($-$0.16, see inset) than for the entire high$-$z sample, indicating the importance of the highest and lowest redshift ranges in the determination of significant metallicity evolution. {\bf Top right panel:} Magellan sample combined with only the highest and lowest redshift bins of the \hiz\ sample. {\bf Bottom left panel:} Magellan sample and only the highest redshift bin of the \hiz\ sample.  {\bf Bottom right panel:} The Magellan and the \hiz\ sample combined. 
}
\label{fig:compare}
\end{figure*}

Following these works, we calculate the cosmological mean metallicity defined as, $\langle Z \rangle = \log [\sum_i 10^{[M/H]_i} N(\hi ) _i / \sum_i N(\hi ) _i] $ where $i$ represents the \dla\ in a given redshift bin. In order to roughly match the bin sizes of R12, we divide the Magellan sample into four redshift bins containing an equal number of \dlas\ (25 \dlas /bin), except for the lowest  redshift bin that contains only 24 \dlas .  We calculate  $\langle Z \rangle$ and plot the results in Figure~\ref{fig:mvz_cosmomean} in red with 1$\sigma$ confidence bootstrap error bars.  As explained in R12, because the $\langle Z \rangle$ is dominated by sample variance rather than statistical error, the bootstrap method best describes a realistic uncertainty.  For the sake of comparison, we have applied the same bootstrap error method as that described in R12.  Overplotted as a grey dashed line is the linear least-squares fit through the $\langle Z \rangle$ values and their uncertainties, described by  $\langle Z \rangle$ = (\fitcosmob )$z -$(1.06$\pm$0.36)%\fitcosmoy
 .  Perhaps surprisingly, this slope is consistent with no redshift evolution in metallicity over the redshift range of the Magellan sample, $z = 2.21 - 4.40$ and possibly in contrast with the results of R12, who found $\langle Z \rangle$ = ($-$0.22$\pm$0.03)$z -$(0.65$\pm$0.09) over the range $z$ = 0.09 $-$ 5.06.
Given the slope of the Magellan sample, in seeming contrast with the previously published surveys, we first consider in greater detail whether the difference in detected evolution between the Magellan sample presented here and the sample of R12 is significant and then discuss the possible effects of potential biases.

\subsection{How different are the Magellan and \hiz\ samples?}\label{sec:howdiff}
One potential source of confusion in comparing the evolution implied by the Magellan and the \hiz\ samples could be attributed to the difference in redshift ranges, with the Magellan sample covering $z = 2.21 - 4.40$ and the \hiz\ sample covering the larger range of $z = 0.09 - 5.06$.  For example, if we constrain the \hiz\ sample to include \emph{only} the data within the redshift range of the Magellan sample, we obtain a best-fit result, $\langle Z \rangle$ = ($-$0.16$\pm$0.07)$z -$(0.82$\pm$0.22), see the top left panel of Figure~\ref{fig:compare}.  This `flattening' of the slope of the \hiz\ sample -- from $-$0.22 to $-$0.16 --  when excluding the \dlas\ in the highest and lowest redshift ranges indicates the significant contribution of these bins to the detection of evolution.  If we instead restrict the redshift range to $z = 2.2 - 3.5$, where the majority of the Magellan sample lies, we find an identical result, $\langle Z \rangle$ = ($-$0.16$\pm$0.12)$z -$(0.82$\pm$0.35).

We have performed the opposite test and calculated the effect on the evolution of cosmic metallicity of combining the Magellan sample with the highest and lowest redshift bins of the \hiz\ sample.  In the top right panel of Figure~\ref{fig:compare} we plot the results of this test, which give $\langle Z \rangle$ = ($-$0.21$\pm$0.03)$z -$(0.63$\pm$0.08), similar to the slope determined from the \hiz\ sample.  This similarity again emphasizes the importance of the highest and lowest redshift bins in measuring metallicity evolution.  Indeed, the slope obtained from \emph{just} the highest and lowest redshift bins of the \hiz\ sample is $\langle Z \rangle$ = ($-$0.23$\pm$0.03)$z -$(0.59$\pm$0.08).

However, as R12 notes, the lowest redshift bin ($z \sim 0 - 1.5$) included in the \hiz\ sample presents a problem because the \dlas\ in this bin were selected based upon strong MgII absorption, typically associated with high metallicity, and therefore constitute a not-unbiased representation of the low redshift end.  Given this fact, we repeat the above test, this time excluding the lowest redshift bin.  In the bottom left panel of Figure~\ref{fig:compare} we plot the metallicity evolution derived from the Magellan sample -- which alone has a slope of $-$0.04 -- and \emph{only} the highest redshift bin from the \hiz\ sample.  The resulting evolution, $\langle Z \rangle$ = ($-$0.23$\pm$0.05)$z -$(0.47$\pm$0.16), even neglecting the lowest, metallicity-biased redshift bin, is again similar to that of the \hiz\ sample (slope = $-0.22$).  This result emphasizes the importance of the highest redshift bin in determining an evolution and, assuming this bin is not biased (however, see \S~\ref{sec:hizbias}), leads us to the conclusion that a possible explanation for the apparent lack of evolution found in the Magellan sample is simply an effect of the limited redshift range covered. Finally, in the bottom right panel of Figure~\ref{fig:compare} we calculate the cosmological mean metallicity evolution of both the Magellan and the \hiz\ samples combined.  While combining the samples produces the highest significance simply from including the most \dlas , we note that caution must be employed in interpreting this result, as shown above and in \S~\ref{sec:bias}, the Magellan sample is relatively unbiased with respect to the \hiz\ sample and it is likely that there are significant biases in this result.  %However, we note that the Magellan sample is relatively unbiased with respect to the \hiz\ sample and it is likely that there are significant biases in this result.  

\begin{figure}
\includegraphics[width=0.99\columnwidth]{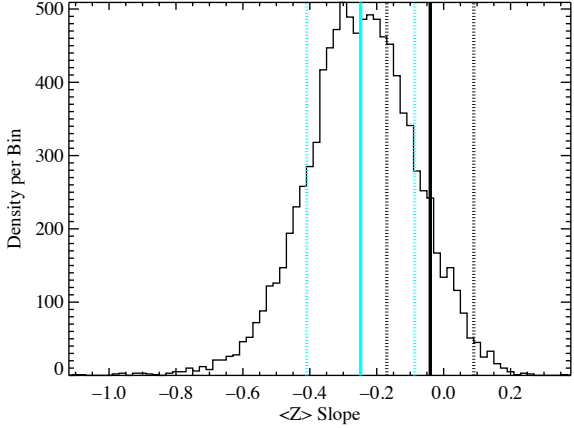}
\caption{Distribution of slopes of the evolution of $\langle$Z$\rangle$ derived from 10,000 bootstrap samples of the \hiz\ metallicity distribution according to the redshift distribution of the Magellan sample.  It is seen that the slope of the \hiz\ sample (mean =  $-$0.25 $\pm$ 0.16, shown in cyan), when taken over the redshift range of the Magellan sample, is, considering error bars (dotted lines), strictly consistent with that of the Magellan sample, shown here in black at \fitcosmob .  Given the \hiz\ sample with the redshift distribution of the Magellan sample, there is a $\sim$10\% probability of obtaining a slope equal to or flatter than the Magellan sample.     
}
\label{fig:bootdist}
\end{figure}

In an attempt to minimize the effects of the redshift range differences of the samples, 
we performed a bootstrap analysis in which we selected \dlas\ from the \hiz\ sample according to the redshift distribution of the Magellan sample, and then calculated the resulting cosmological means and best-fit slope.  We repeated this process 10,000 times, always randomly selecting the \hiz\ sample \dlas\ with replacement according to the redshift distribution of the Magellan sample.  In Figure~\ref{fig:bootdist} we plot a histogram of the resultant slopes.  The mean slope, plotted as a cyan vertical line, is $-0.25$, with a standard deviation of 0.17 (dotted cyan lines).  We compare this with the slope of the Magellan sample, \fitcosmob , shown as a vertical black line.  It is clear that the bootstrap distribution from the R12 sample overlaps considerably with the error margin from the Magellan sample. Indeed, there is a $\sim$10\% chance of obtaining a slope equal to the Magellan sample ($-$0.04) or flatter if the R12 sample is selected according to the redshift distribution of the Magellan sample. In other words, after considering the differences in their redshift distributions, the lack of metallicity evolution found in the Magellan sample is not in conflict with that of the R12 sample.

%there is a not insignificant ($\approx$10\%) chance to obtain a slope equal to the Magellan sample ($-$0.04) or flatter, given the \hiz\ sample distributed according the redshift distribution of the Magellan sample.  In other words, the evolution found by the Magellan sample is not in conflict with that of the \hiz\ sample.  

%%F12

%%F13
\begin{figure}
\includegraphics[width=0.99\columnwidth]{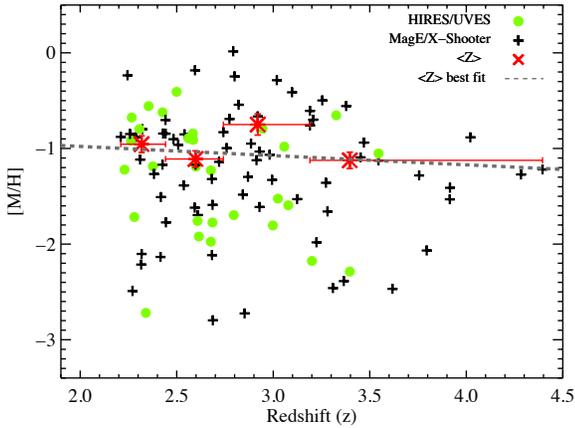}
\caption{
[M/H] versus redshift for the flux-corrected Magellan sample, assuming b = 10 km s$^{-1}$.  The sample is divided into four redshift bins containing equal numbers of \dlas .  The red points denote the cosmic mean metallicity with 1$\sigma$ confidence interval bootstrap error bars as defined in the text. The grey dashed lines represent the linear best-fit to the binned data points, $\langle Z \rangle$ = ($-$0.10 $\pm$ 0.11 )$z -$(0.79 $\pm$ 0.31).
}
\label{fig:mvz_cosmomean_b10}
\end{figure}

\subsection{Assessment of Potential Biases}\label{sec:bias}
\subsubsection{Bias in the Magellan sample?}\label{sec:lozbias}
We first note that the detection of $\sim$zero evolution in the Magellan sample is not dependent on the highest redshift \dlas\ -- in fact, if we exclude the three objects at z$_{abs} >$4 we derived a similar result, $\langle Z \rangle$ = ($-$0.01$\pm$0.13)$z -$(1.15$\pm$0.38).  We do note however, that if we remove the three lowest metallicity \dlas\ from the Magellan sample, all with [M/H]$<$-2.7, we obtain $\langle Z \rangle$ = ($-$0.10$\pm$0.12)$z -$(0.87$\pm$0.32).  While this is still technically consistent with no evolution, it is interesting that the removal of just three (low metallicity) \dlas\ moves the slope in a direction consistent with the \hiz\ sample.  Part of this change in slope is caused by a slight change in binning due to the removal of three \dlas .  Specifically, the high metallicity, high \nhi\  \dla\ 1344$-$0323, is moved from the highest redshift bin to the neighboring bin, creating some of the change in slope.  However, we note that the metallicity values of the three low metallicity \dlas\ are relatively secure: one object, \dla\ 0035$-$0918, is taken from Keck/HIRES data ~\citep{cooke11}, while the other two, \dla\ 1337$-$0246 and \dla\ 1358$+$0349 are supported by expected/similar [Fe/H] measurements, i.e. the [Fe/H] values are similar and less than 0.3 dex different than the [$\alpha$/H] value as expected for low-metallicity \dlas , i.e. Figure 11 of R12.  Is it by chance that the Magellan sample contains three relatively low metallicity \dlas\ at z$_{abs} \sim$2.5?  We discuss this question further in the next section, \S~\ref{sec:hizbias}.

Although we have already shown in ~\S~\ref{sec:measure} that by applying a minimum flux cut (here, F$_{min}$/F$_q <$ 0.65), additional flux-based column density corrections will likely not have a large effect on the overall measurements, we did investigate the effects of applying the flux-based correction on the evolution of $\langle Z \rangle$.  In Figure~\ref{fig:mvz_cosmomean_b10} we plot the results of applying the flux-based column density correction shown in Figure~\ref{fig:deltan} for a Doppler parameter b = 10 \kms\ (these corrections are also shown in Figure~\ref{fig:mvz_corrected}).  It is seen that this does have some effect on the slope, with a best-fit of  $\langle Z \rangle$ = ($-$0.10$\pm$0.11)$z -$(0.79$\pm$0.31).  Interestingly, the application of the flux-based corrections does move the measured evolution in $\langle Z \rangle$ closer to the results of R12.  However, while the change is insufficient for full agreement we cannot rule out that this could be a contributing effect to the detected difference in the $\langle Z \rangle$ evolution of the Magellan and \hiz\ samples. 

\subsubsection{Bias in the \hiz\ Sample?}\label{sec:hizbias}

 In total, the \hiz\ sample contains 33 \dlas\ with medium-resolution Keck/ESI spectra only, 22 \dlas\ with Keck/HIRES spectroscopy, and an additional 195 \dlas\ taken from the literature. For their newly presented metallicities, R12 state that they did not apply any flux corrections to their medium-resolution Keck/ESI data because they followed-up all likely problematic candidates with Keck/HIRES in order to obtain a good measurement of the metallicity. 

 %%F14
\begin{figure*}
\includegraphics[width=1.75\columnwidth]{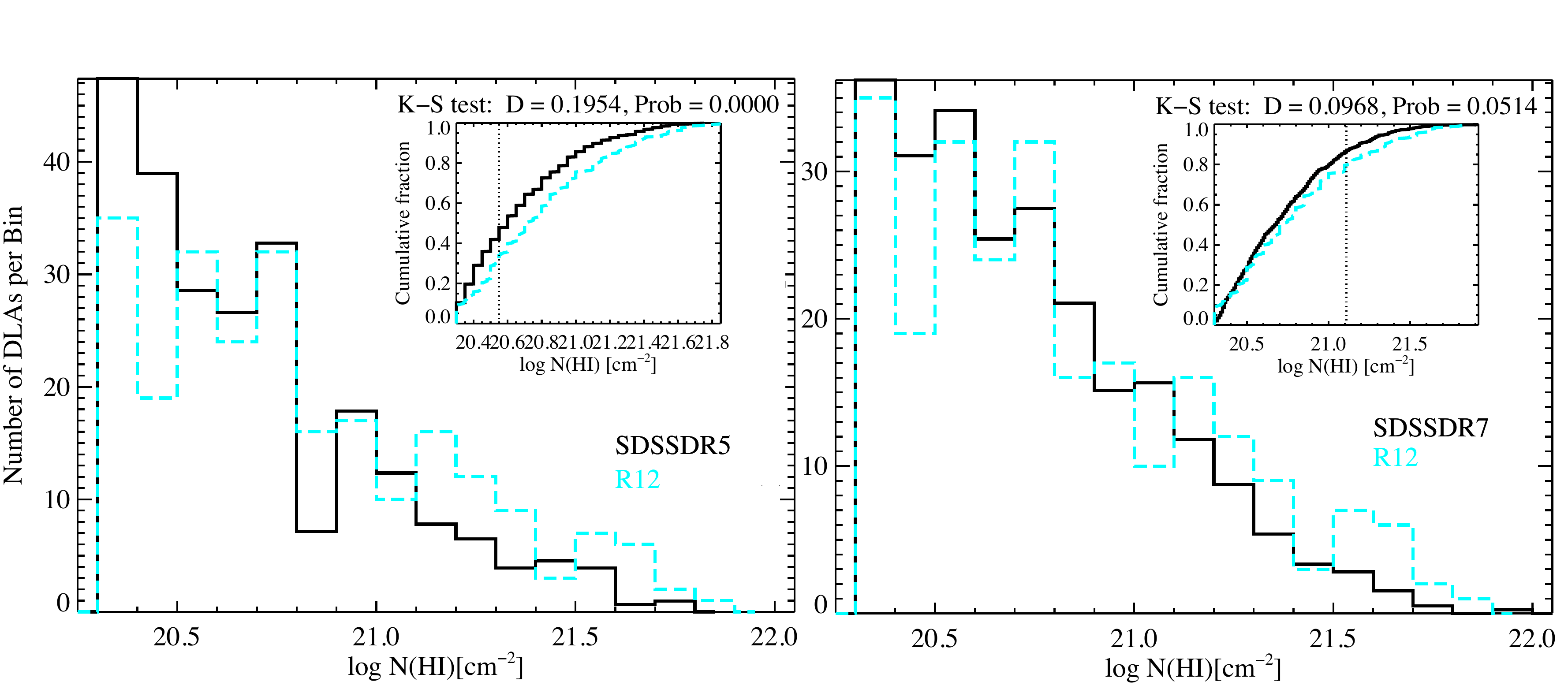}\\
\caption{\nhi\ histogram comparing the \hiz\ sample (cyan, dashed line) with the scaled SDSSDR5 ~\citep{pro05} distribution, left, and the scaled SDSSDR7 ~\citep{noterdaeme09} distribution, right.  A K-S test indicates that the probability they are drawn from the same parent population is P$_{K-S}$ = 1 $\times$10$^{-6}$ and P$_{K-S}$ = 0.05, respectively. 
}
\label{fig:marcvdr7}
\end{figure*}

 In contrast to the Magellan sample that was specifically designed \emph{a priori} to be uniformly selected, as demonstrated by the good agreement in \hi\ column density distribution, f(\nhi ), between the Magellan sample and the SDSS sample (i.e. Figure~\ref{fig:h2hist}), the \hiz\ sample contains a large number ($\sim$195) of \dlas\ taken from the literature.  While R12 state that care was taken to avoid including \dlas\ from biased samples, it is still illuminating to compare the \hi\ column density distribution of the \hiz\ sample with that of the SDSS to assess any potential level of bias.  We show the results of this comparison in Figure~\ref{fig:marcvdr7}, where a two sided Kolmogorov-Smirnov (K-S) test shows that the probability of the \hiz\ sample and the SDSS DR5 (left) and SDSS DR7 (right) sample to be drawn from the same parent population is P$_{KS}$ = 1$\times$10$^{-6}$, and P$_{KS}$ = 0.05 respectively.  It is seen that there may be a slight deficit of low-\nhi\ systems and overabundance at the high-\nhi\ end.

Indeed, if we assume that the metallicity distribution of the \hiz\ sample is an unbiased representation of the true \dla\ metallicity distribution, we should be able to compare it with that of the Magellan sample. In order to facilitate comparison between the samples we consider only those \dla\ within the redshift range of the Magellan sample and fit the metallicity distribution with a Gaussian, as shown in Figure 8 of R12.  We derive a best-fit Gaussian with mean metallicity [M/H] = $-$1.54 and $\sigma$=0.46. Assuming this distribution is correct,  we would expect less than $\sim$0.5\% of any sample to have [M/H]$\leq\  -2.72$.  The Magellan sample contains 3/99 \dlas , or $\sim$3\% of the sample with [M/H]$\leq  -2.72$, an interesting, but perhaps not significant difference from the distribution expected from the \hiz\ sample.

As previously mentioned, an additional potentially large bias in the \hiz\ sample, also discussed by R12, is the inclusion of the \dlas\ in the lowest redshift bin ($z \sim\ 0 - 1.5$).  Because these DLAs were generally first identified by their strong MgII absorption, it would be unsurprising to find they are biased towards higher metallicities.

\subsection{Conclusions on Metallicity Evolution}

The previously detected cosmic mean metallicity evolution derived from \dlas\ was measured by R12, to be $\langle Z \rangle$ = ($-$0.22$\pm$0.03)$z -$(0.65$\pm$0.09) over the range $z$ = 0.09 $-$ 5.06.  In this paper, we present an independent, albeit smaller \dla\ sample that found essentially no evolution, $\langle Z \rangle$ = (\fitcosmob )$z -$(1.06$\pm$0.36) over the redshift range $z$ = 2.21 $-$ 4.40.  We note that the majority of this sample falls between $z$ = 2.21 $-$ 3.50, and that the slopes of the R12 and Magellan samples are, strictly speaking, consistent with each other at the 2$\sigma$ level.

%We emphasize several points about this result: First, the majority of the Magellan sample presented here falls between $z$ = 2.21 $-$ 3.50, a smaller redshift range than that of R12, and second, the larger error bars of the smaller Magellan sample mean that the slopes of the R12 and Magellan samples are, strictly speaking, consistent with each other at the 2$\sigma$ level.  

As discussed in \S~\ref{sec:howdiff} much of the power of the detected evolution in the \hiz\ sample comes from the highest and lowest redshift bins covering a redshift space not probed by the Magellan sample.  While this fact and the biases outlined in \S~\ref{sec:lozbias} and \S~\ref{sec:hizbias} may prohibit a direct comparison of the measurements of cosmic metallicity evolution (or lack thereof), a relevant question remains:  Is there evolution in metallicity over the redshift range probed by the Magellan sample, $z \sim\ 2 - 4$?

In answering this question, we consider the following facts:  1) The Magellan sample was designed \emph{a priori} to be uniformly selected and is more consistent with the \nhi\ frequency distribution function of the parent SDSS sample,   2) the slope found by the \hiz\ sample is heavily weighted by the lowest and highest redshift ends, and 3) the bootstrapping of the \hiz\ sample within the redshift range of the Magellan sample indicates that the samples are not inconsistent with each other.  Given this evidence, we propose that the slope of metallicity evolution of \dlas\ between $z \sim\ 2 - 4$ may be flatter than that found by the \hiz\ sample and closer to the most likely value found in the Magellan sample presented here.

\section{Other \dla\ Diagnostics}\label{sec:others}
In this section we present an analysis of several additional \dla\ diagnostics that are important in determining \dla\ gas properties such as the width of the low-ion velocity profile, \delvninty , and the equivalent width of the Si II $\lambda$1526\AA\ line, \ewsitwo .

\subsection{$\Delta v_{90}$}\label{sec:delv}

%%F15
\begin{figure*}
\includegraphics[width=1.75\columnwidth]{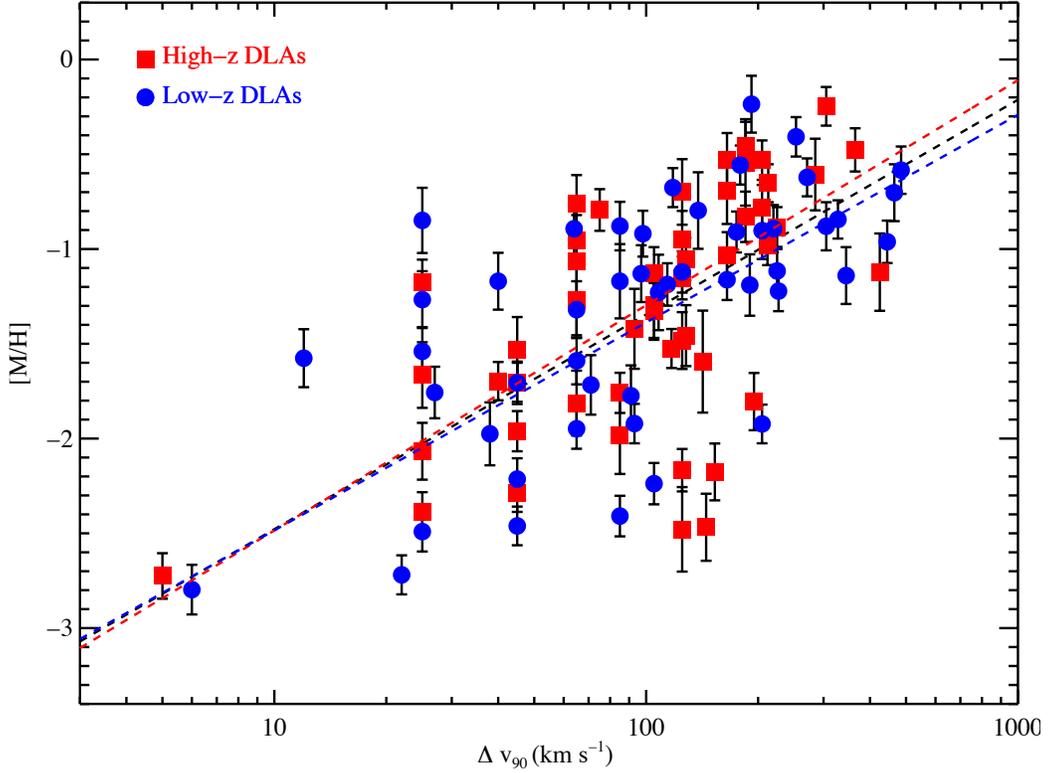}\\
\caption{Metallicity versus $\Delta$v$_{90}$, a measure of the kinematic state of the gas, for the Magellan sample.  Red squares denote the high-$z$ bin with all points greater than or equal to the median z$_{abs}$ = \zmedian , while blue circles denote the low$-z$ bin.  While there does appear to be a correlation between $\Delta$v$_{90}$ and [M/H], there is no evidence for evolution with redshift, as seen by the similarity of the red and blue distributions. The dashed lines represent power law fits as described in the text for the entire sample (black), the high-$z$ sample (red) and the low$-z$ sample (blue).  
}
\label{fig:delvvm}
\end{figure*}

The $\Delta v_{90}$ statistic is defined to be a measure of the velocity interval that contains 90\% of the integrated optical depth of the low-ion metallic gas ~\citep{pro97}.  \delvninty\ is typically measured from an unsaturated low-ion transition and represents the kinematic state of the bulk of the gas.  In past \dla\ surveys several authors, i.e.  ~\cite{ledoux06} and ~\cite{pro08}, have shown this statistic to be strongly correlated with \dla\ metallicity.  They interpret this as a sort of mass--metallicity relation and derive slopes that are similar to those found for the mass--metallicity relationships in samples of local, low-metallicity galaxies.

Following the practice of ~\cite{pro08}, who analyze the effects of lower resolution spectra in determining the $\Delta v_{90}$ parameter and find that for their medium-resolution Keck/ESI spectra, the $\Delta v_{90}$ values are biased high by approximately half of the instrumental FWHM, we assume that our MagE spectra (with FWHM $\sim$ 71 \kms ) are biased high by $\sim$ 35 km s$^{-1}$.  Therefore, we reduce the $\Delta v_{90}$ values obtained from the MagE data by 35 \kms , in order to account for this systematic effect. Likewise, for the X-Shooter data (FWHM $\approx$ 59 \kms ) we reduce the $\Delta v_{90}$ values by 30 km s$^{-1}$.  We report the \delvninty\ statistic in column 5 of Table~\ref{tab:sample}.

In Figure~\ref{fig:delvvm} we plot [M/H] versus \delvninty\ for the Magellan sample. It can be seen by eye that there is a correlation, albeit with $\sim$1.5\,dex scatter, between \delvninty\ and \dla\ metallicity.  As stated by others ~\citep{ledoux06, pro08}, this scatter is likely due to differences in impact parameter and the inclination of the galaxy over which \delvninty\ is measured.  Statistically, the correlation is significant -- the Kendall tau rank correlation provides a probability P($\tau$) = 9.6 $\times$ 10$^{-11}$ of being due to chance alone, corresponding to significance of correlation $>$6.4$\sigma$. %We find a linear Pearson correlation coefficient of r = \linpdelvm .  
This result is in agreement with previous surveys that have found similar trends, i.e. ~\cite{ledoux06, pro08}.

A power law fit to the $\Delta v_{90}$ vs.~[M/H] data in Figure~\ref{fig:delvvm},

\begin{equation}
[M/H] = a + b\ \mathrm{ log}( \Delta v_{90})
\end{equation}      

\noindent drawn as the black dashed line, gives best fit parameters a = \delvvma\ and b = \delvvmb .  This is a somewhat flatter slope than that found by ~\cite{ledoux06}, who report,  a = $-$4.33 $\pm$ 0.23 and b = 1.55 $\pm$ 0.12.  While we hesitate to speculate in depth on the nature of this discrepancy, we point out several differences between the ~\cite{ledoux06} sample and the Magellan sample: The \cite{ledoux06} sample 1) is smaller (70 objects), 2) includes 13 Super Lyman-limit Systems (SLLS), which could introduce additional confusion from ionization corrections, and 3) is known to have an f(\nhi ) distribution different than the `unbiased' SDSS distribution. 

 %%F16
\begin{figure*}
\includegraphics[width=1.75\columnwidth]{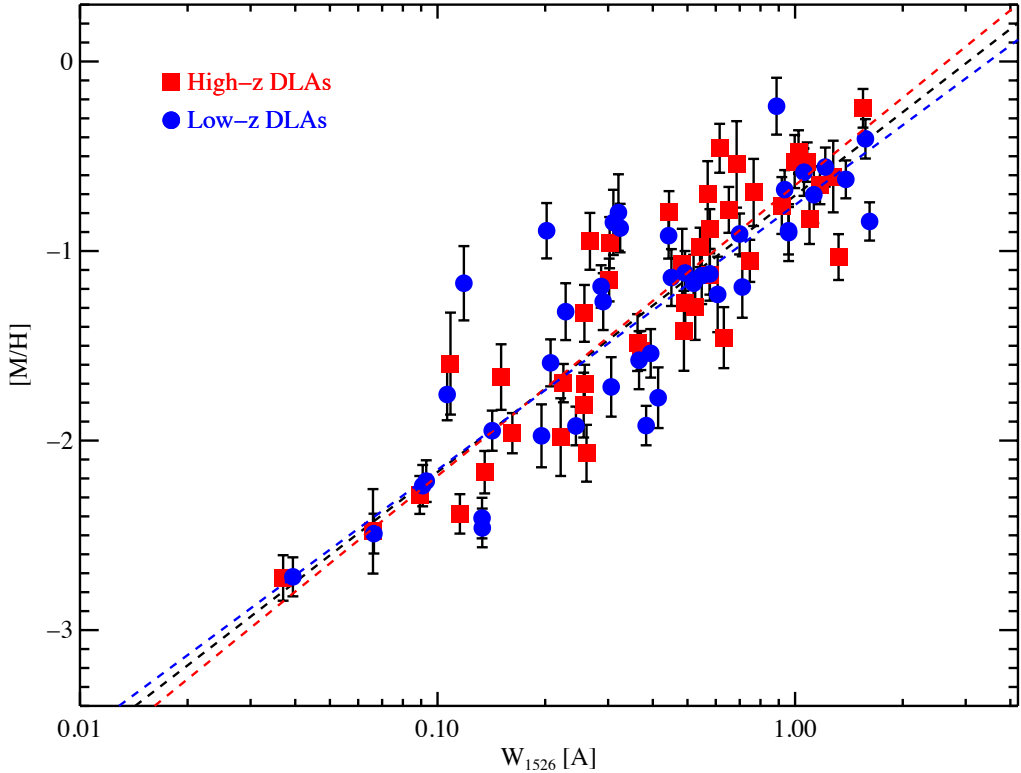}\\
\caption{The remarkably tight correlation between Si II 1526\AA\ Equivalent Width versus metallicity in the Magellan sample.  Red points denote the high-$z$ bin with all points greater than or equal to the median z$_{abs}$ = \zmedian , while blue points denote the low$-z$ bin.  It is visually apparent that there is essentially no significant evolution with redshift.  The dashed lines represent power law fits as described in the text.  The average error on the W$_{1526}$ values is $\sim$ 0.013 \AA 
}
\label{fig:sivm}
\end{figure*}

The median \delvninty\ of the Magellan sample is \delvvmmed\ km s$^{-1}$.  This is higher than the median \delvninty\ found by ~\cite{pro08}, 80 \kms , and the medians found by ~\cite{ledoux06} in their high and low redshift samples, 69 and 92 \kms , respectively. 
To investigate the possibility of redshift evolution in the \delvninty\ parameter, as found by ~\cite{ledoux06}, we split the sample into high and low redshift bins, separated by the median z$_{abs}$ = \zmedian .  We find, in contrast to the results of ~\cite{ledoux06}, that there is little difference between these two populations.  The median metallicity and median velocity width of the two populations are $\langle $ [M/H] $^{high z}$ $\rangle $= \medianmtlhi\ and$\langle $ \delvninty $^{high z}$ $\rangle $ =  \mediandelvhi\ in the high redshift bin, and $\langle $ [M/H]$^{low z}$ $\rangle $ = \medianmtllo\ and $\langle $ \delvninty $^{low z}$ $\rangle $ =  \mediandelvlo , in the low redshift bin. While the metallicity is virtually unchanged, there is a slight, yet insignificant decrease in \delvninty\ from high to low redshift. Strictly speaking, this is opposite to the trend found by ~\cite{ledoux06}, however, given the error bars it is likely not significant. This can be seen in Figure~\ref{fig:delvvm} where the red squares (blue circles) indicate the high (low) redshift bin.  It is seen by eye that there is very little difference in the two distributions.  A two sided Kolmogorov-Smirnov (K-S) test shows that the low- and high-$z$ \delvninty\ values are entirely consistent with being drawn from the same parent population, P$_{KS}$ = \delvzp .

Performing a linear least squares fit to the high and low redshift samples separately gives the following best-fit parameters:
 a$^{highz}$ = \delvvmahi\ and b$^{highz}$ = \delvvmbhi\ ;  
 a$^{lowz}$ = \delvvmalo\ and b$^{lowz}$ = \delvvmblo . 
 The fits are represented by the red (high-$z$) and blue (low$-z$) dashed lines in Figure~\ref{fig:delvvm} and show very little difference between the high and low redshift samples with the possible exception of a slight steepening of the high redshift slope.

\subsection{\ewsitwo\  and \ewciv\ }
 
Another kinematic diagnostic often used to characterize the \dla\ population is the rest equivalent width, defined as W = W$_{obs}$/(1 $+$ z), of various absorption lines.  As explained in ~\cite{pro08} the W statistic is a measure of the kinematics of the system when the line used is optically thick. We report the rest equivalent widths of two transitions representing the low-ion and high-ion transitions, respectively, Si II $\lambda$1526 and C IV $\lambda$1548 in Table~\ref{tab:sample}.   For reference, the expected \ewsitwo\ of the Si II $\lambda$1526 line to become optically thick ($\tau\ > 1$) is $\approx 0.1-0.3$ \AA , depending upon the velocity profile of the system.  

We plot the results of \ewsitwo\  versus metallicity in Figure~\ref{fig:sivm}. We include only \dlas\ with good spectral coverage of the Si II $\lambda$1526 line. If there was no coverage, or if the line suffered from serious blending with an interloper or forest line, we did not include it in this analysis. Figure~\ref{fig:sivm} contains a total of 86 \dlas .

Interestingly, as found by previous authors, the correlation between \ewsitwo\  and metallicity is significantly stronger than that between \delvninty\ and metallicity.  The Pearson correlation coefficient is r = \linpsi , as compared with r = \linpdelvm\ for [M/H] with \delvninty .  The results of the Kendall tau test give a probability that there is no underlying correlation and that the observed $r$ occurs by chance alone, P($\tau$) = 3.9 $\times$ 10$^{-19}$, corresponding to significance of correlation $>$9$\sigma$. Perhaps in part this is not surprising given that the equivalent width of a line should be independent of spectral resolution, as opposed to \delvninty\ which is clearly affected by instrumental FWHM. %, and hence \ewsitwo\  is more closely predictive of the underlying physical truth.  
However, it is still surprising how tight the correlation is given that our data span the optically-thin/optically-thick transition, yet the correlation remains tight and strong everywhere.  A power law fit to the data, 

\begin{equation}\label{eq:fit}
[M/H] = a + b\ \mathrm{ log}(W/(1\AA ))
\end{equation}      
     
\noindent results in best-fit parameters a = \sivma\ and b = \sivmb\ and is denoted by the black dashed line.  While the results for the slope of the Magellan sample agree well with those of ~\cite{pro08} who derived best-fit parameters for their sample, a = $-$0.92 $\pm$ 0.05, b = 1.41 $\pm$ 0.10, the y-intercepts are different by 0.05 implying that metallicities in the Magellan sample are generally higher at a given W.  

As discussed in ~\cite{pro08}, the fact that the correlation between [M/H] and \ewsitwo\  is even tighter than that between [M/H] and \delvninty\ is perhaps surprising given that the gas that determines the [M/H] -- the bulk of the ISM gas measured in the low-ion components -- is in general physically more related to the \delvninty\ statistic that is also derived from this low-ion gas.  On the contrary, the \ewsitwo\  statistic can contain a large contribution from halo gas unrelated to the bulk of the galaxy.  Hence, as ~\cite{pro08} point out, this implies a mysterious connection between the local ISM properties of the galaxy that determine [M/H] and its environment, specifically, its large scale velocity field, and has been interpreted as a mass-metallicity relation. 

 %We argue here that perhaps the larger spread in the \delvninty\ versus [M/H] relation, as compared with \ewsitwo\ versus [M/H], is actually evidence {\it supporting} the disk-like \dla\ (or some kind of organized motion) scenario.  Specifically, if \dlas\ are disk-like structures (or structures participating in some type of organized motion) one would expect the measured \delvninty\ to depend on the disk inclination angle.  Given a random range of inclination angles, one might expect a spread in relations involving the \delvninty\ statistic.  On the other hand, because  %Apart from adding to the evidence for it, we can offer no further insight into this apparent mystery at this stage.  

Taking this interpretation one step further, we propose that the large difference in scatter of the two correlations is actually evidence {\it supporting} the disk-like \dla\ scenario.  Specifically, because \ewsitwo\ is dominated by gas at large velocities, in the outskirts or halo of the galaxy, a natural interpretation of this correlation is a mass-metallicity relation where the more massive halos are more metal enriched and contain gas at larger velocities that would contribute to the \ewsitwo .  In this case, because the dark matter halo is essentially spherical, the inclination angle of the disk does not matter.  On the other hand, the \delvninty\ parameter is more susceptible to the orientation -- assuming \dlas\ are disk-like structures participating in some organized motion.  In this case, the measured \delvninty\ will depend on the impact parameter and inclination angle of the disk, which would vary widely amongst \dlas\ and provide the source of the larger scatter in the \delvninty\ - [M/H] relation.  Therefore, in some sense, the larger scatter in the \delvninty\  - [M/H] correlation along with a simultaneously smaller scatter in the \ewsitwo\ - [M/H] correlation is consistent with the interpretation of \dlas\ as disk-like structures.  On the other hand, if \dlas\ were primarily merging clumps of gas ~\citep{1998ApJ...495..647H}, it would be difficult to explain the tighter  \ewsitwo\ - [M/H] correlation.

%Specifically, if \ewsitwo\ is dominated by the gas in the outskirts, or the halo of the galaxy, a natural link to a mass-metallicity relation is realized, where the more massive halos are more metal enriched and contain gas at larger velocities that would contribute to the \ewsitwo .  On the other hand, the \delvninty\ parameter is more susceptible to the orientation -- assuming \dlas\ are disk-like structures participating in some organized motion.  In this case, the measured \delvninty\ will depend on the impact parameter and inclination angle of the disk, which would vary widely amongst \dlas\ and provide the source of the larger scatter in the \delvninty\ versus [M/H] relation.  Therefore, in some sense, this large scatter in \delvninty\ versus metallicity and small scatter in \ewsitwo\ versus metallicity supports the idea that \dlas\ are in fact disk like structures, because if they were only random clumps of merging gas, the \ewsitwo\ relation would not be so tight.  

%%F17
\begin{figure}
\includegraphics[width=0.99\columnwidth]{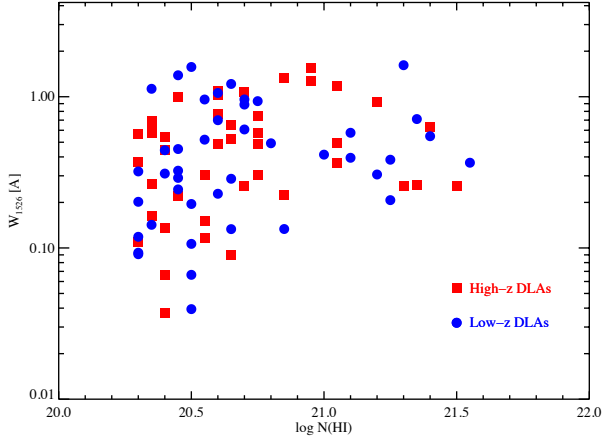}
\caption{Log \nhi\ versus Si II 1526\AA\ Equivalent Width. The average error on the W$_{1526}$ values is $\sim$ 0.013 \AA , while the typical errors on the log \nhi\ values are $\pm$ 0.1 dex . Red points denote the high-$z$ bin with all points greater than or equal to the median z$_{abs}$ = \zmedian , while blue points denote the low$-z$ bin.  
}
\label{fig:nhivsii}
\end{figure}

Dividing the \ewsitwo\ sample into two bins of high and low redshift reveals no evidence for any evolution in the \ewsitwo\ parameter with redshift.  The high redshift median $\langle $ \ewsitwo\ $^{high z} \rangle $ = \ewhi\ while the low redshift median $\langle $ \ewsitwo\ $^{low z}$ $\rangle $ = \ewlo .  We note that, if \ewsitwo\ really is a good tracer of metallicity, then these medians imply a \emph {decrease} in metallicity with decreasing redshift, contrary to what one would expect if \dlas\ are indeed tracing the build-up of metals over cosmic time.  Interestingly, this is the same behavior as seen in \delvninty , where the median value of the \delvninty\ parameter actually decreases slightly with redshift -- an opposite trend to that reported in \delvninty\ by ~\cite{ledoux06}.  However, we note that within the error bars, there is no significant change in the median values.  Again, this can be seen in Figure~\ref{fig:sivm} where we have plotted the high (low) redshift points as red squares (blue circles).  The two populations virtually overlap.
 Best fit lines to the data produce the dashed red and blue lines for the high and low redshift bins respectively.  The best-fit linear parameters are 
 a$^{highz}$ = \sivmahi\ and b$^{highz}$ = \sivmbhi ;  a$^{lowz}$ = \sivmalo\ and b$^{lowz}$ = \sivmblo .  Any difference in the low and high redshift populations is not very significant as the two sided Kolmogorov-Smirnov test gives the probability that the two are drawn from the same parent population as P$_{KS}$ = \ewpks . 

%%F18
\begin{figure}
\includegraphics[width=0.99\columnwidth]{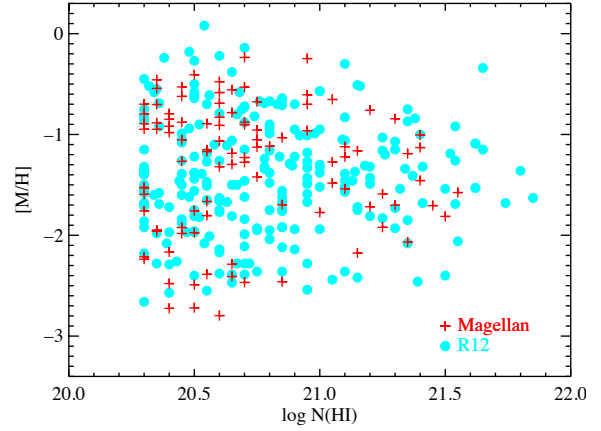}
\caption{Log \nhi\ versus [M/H] for the Magellan sample (red crosses) and the \hiz\ sample (cyan circles).  It is clear that the correlation between log \nhi\ and W$_{1526}$ at low W$_{1526}$ seen in Figure~\ref{fig:nhivsii}, is related to a similar trend seen here with [M/H] in the Magellan sample.  However, the \hiz\ sample (cyan circles) does not seem to show this same trend, indicating that perhaps rather than having a physical origin, it is the result of small number statistics.
}
\label{fig:nhivmet}
\end{figure}

%%F19
\begin{figure}
\includegraphics[width=0.99\columnwidth]{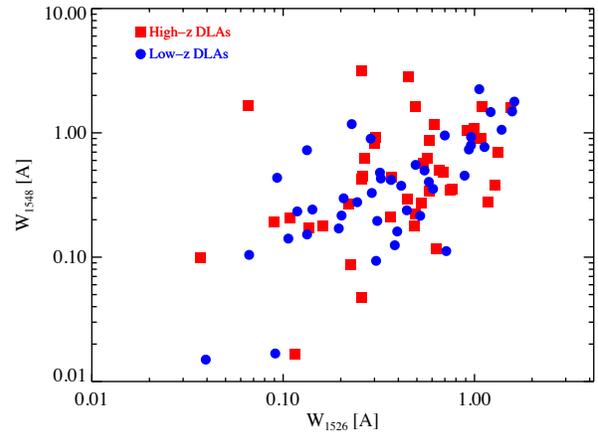}
\caption{C IV 1548\AA\ Equivalent Width versus Si II 1526\AA\ Equivalent Width of the Magellan sample.  Red points denote the high-$z$ bin with all points greater than or equal to the median z$_{abs}$ = \zmedian , while blue points denote the low$-z$ bin. The average error on the W$_{1526}$ and W$_{1548}$ values is $\sim$ 0.013 \AA\ and $\sim$ 0.015 \AA , respectively.
}
\label{fig:siivciv}
\end{figure}

We also examine \ewsitwo\  as a function of \nhi , as seen in Figure~\ref{fig:nhivsii}.  Similar to ~\cite{pro08} (Figure 7) we see a trend for increasing N(\hi ) with increasing \ewsitwo\ up to \ewsitwo\ $\sim$ 0.2 \AA , where the data then become a scatter plot.  The results of the Kendall tau test give a significance of correlation $>$2.6$\sigma$ (P($\tau$) = 9.5 $\times$ 10$^{-3}$).  One immediate question raised by Figure~\ref{fig:nhivsii} is why there are no \dlas\ with low \ewsitwo ($\sim \le$ 0.2 \AA ) and high \nhi ? Is there some physical mechanism to explain this apparent `forbidden zone?'

In looking at the behavior of the metallicity of the Magellan sample as a function of \nhi\ (see Figure~\ref{fig:nhivmet}, red crosses), we see a similar `forbidden zone' in the region of low metallicity and high \nhi .  That we see this same trend is not surprising, as we have shown above that \ewsitwo\ is tightly correlated with metallicity. However, when we compare this with the \hiz\ sample -- cyan circles in Figure~\ref{fig:nhivmet} -- we see that any type of correlation between \nhi\ and metallicity seems to disappear.  Indeed, the results of the Kendall tau test, P($\tau$) = 0.63, indicate no correlation, leading us to conclude that any correlation seen at low \ewsitwo\ is likely the result of small number statistics. 

A correlation is seen for the relation between \ewsitwo\ and the equivalent width of high-ion gas, \ewciv\ as seen in Figure~\ref{fig:siivciv}.  The linear Pearson correlation coefficient is r=0.43 and a positive correlation between the two variables is detected at the $>$5.5$\sigma$ level (P($\tau$) = 3.2 $\times$ 10$^{-8}$).  A similar correlation was reported by ~\cite{pro08} and has been interpreted as a sign that the low and high-ion gas is subject to a common gravitational potential well ~\citep{wolfe00, maller03}.

%\subsection{\ciistr }

%As shown by ~\cite{wolfe03a}, the heating rate per hydrogen atom in a \dla , and hence with some assumptions, the star formation rate, can be estimated by measuring the \ciistr\ $\lambda$1335 transition.  
%We measured or put an upper limit on N(\ciistr ) in each \dla , and we present the results in Table~\ref{tab:sample} and Figure~\ref{fig:cii}.  We find a total of 18 \ciistr\ detections.  We note that generally high-resolution is required to accurately measure N(\ciistr ) as shown in ~\cite{wolfe08}, and hence we are possibly underestimating the number of \ciistr\ detections in this medium-resolution survey. 

\subsection{Mg II $\lambda$2796 Equivalent Width versus [M/H] and \nhi\ }

\begin{figure}
\includegraphics[width=0.99\columnwidth]{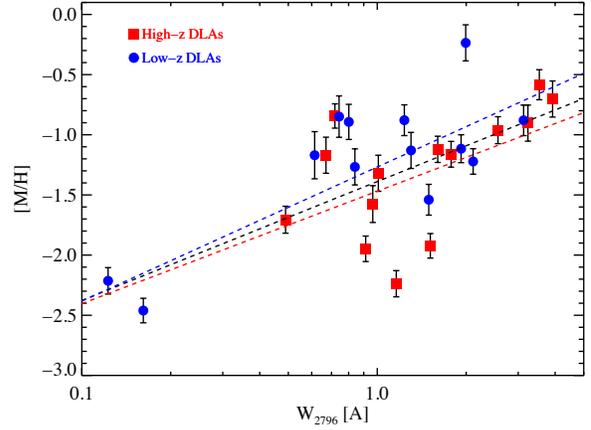}
\caption{[M/H] versus MgII $\lambda$2796\AA\ equivalent width (W$_{2796}$) of the Magellan sample that contained spectral coverage of the MgII $\lambda$2796\AA\ line.  The mean error on W$_{2796}$ is $\sim$ $\pm$ 0.12 \AA .  Red squares denote the high redshift points, while blue circles represent the low redshift points, where the sample was split on the median redshift, z$_{abs}$=2.44.  While the correlation is not highly significant ($\sim$ 2.5$\sigma$), there is a trend for the highest equivalent width systems to also have high metallicities.  There is is also a slight trend for increased metallicities in the lower redshift sample for a given W$_{2796}$.
}
\label{fig:mgvm}
\end{figure}

We find some evidence for a correlation between metallicity and the equivalent width of the MgII $\lambda$ 2796 \AA\ line, W$_{\lambda 2796}$, in the Magellan sample, see Figure~\ref{fig:mgvm}.  %The linear Pearson Correlation Coefficient r = 0.58.   
The results of the Kendall tau test show a significance of correlation $>$2.5$\sigma$ (P($\tau$) = 1.2 $\times$ 10$^{-2}$).  This is a similar trend to that found by ~\cite{murphy07}, who report a 4.2$\sigma$ significant correlation between W$_{\lambda 2796}$ and [M/H] in a sample of 49 \dlas\ and strong sub-\dlas .  ~\cite{murphy07} conclude that this correlation is a result of the connection between an absorber's metallicity and the mechanism for producing and dispersing the velocity components, since the saturated $W_{\lambda 2796}$ is most sensitive to the kinematic spread of the gas.

A power law fit to the data, like that in equation ~\ref{eq:fit}, results in best-fit parameters a = $-$1.38 $\pm$ 0.02 and b = 0.97 $\pm $0.06 and is shown in Figure~\ref{fig:mgvm} by the black dashed line.  Dividing the already small sample (27 \dlas) into high and low redshift bins about the median redshift of this sample, $z = 2.44$, gives best-fit parameters of the power law fits to the subsamples: a$^{highz}$ = $-$1.47 $\pm$ 0.04 and b$^{highz}$ = 0.93 $\pm$ 0.12 and a$^{lowz}$ = $-$1.25 $\pm$ 0.04 and b$^{lowz}$ = 1.10 $\pm$ 0.08.  These fits are shown in Figure~\ref{fig:mgvm} by the red and blue dashed lines, respectively.  While these fits may indicate some evolution in $W_{\lambda 2796}$ with redshift -- indeed the median W$_{\lambda 2796}$ deceases from high to low redshift, $\langle $ W$_{\lambda 2796}$ $^{high z} \rangle $ = 1.51 $\pm$ 0.79 \AA\ to $\langle $ W$_{\lambda 2796}$ $^{low z} \rangle $ = 1.23 $\pm$ 0.62 \AA\ --  given the sizable error bars, this change may not be significant.  %Moreover, the relatively small sample size of just 27 \dlas\ makes this result difficult to interpret further. 

\begin{figure}
\includegraphics[width=0.99\columnwidth]{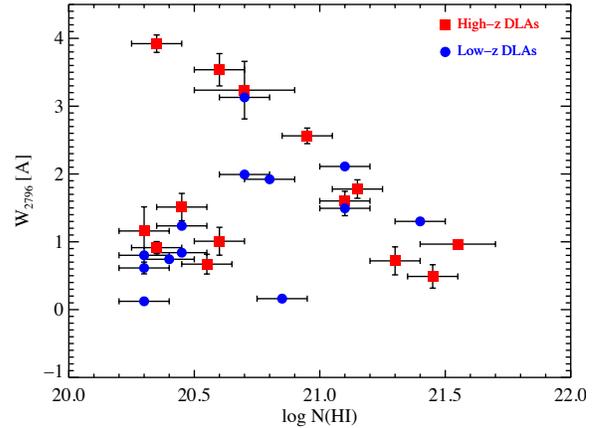}
\caption{MgII $\lambda$2796\AA\ equivalent width versus log N(\hi ) for \dlas\ in the Magellan sample that contained spectral coverage of the MgII $\lambda$2796\AA\ line.  Red squares denote the high redshift points, while blue circles represent the low redshift points, where the sample was split on the median redshift, z$_{abs}$=2.44.
}
\label{fig:mgvnhi}
\end{figure}

We find no evidence for a correlation between $W_{\lambda 2796}$ and \nhi\ as seen in Figure~\ref{fig:mgvnhi}.  The results of the Kendall tau test, P($\tau$) = 0.52, provide essentially no evidence for correlation. The noticeable absence of systems with \emph{both} large log\nhi\ and $W_{\lambda 2796}$, in the upper right corner of the plot, is not statistically significant because of the small number of \dlas\ available with $W_{\lambda2796}$ measurements.

%%F20

%%F21
	
\section{Conclusions}~\label{sec:discussion}

We present the first uniformly selected \dla\ sample with medium-resolution (or higher) measurements of metallicity, \delvninty\ and \ewsitwo .  This sample is unique in the sense that it was created \emph{a priori} to be as unbiased as possible by including all SDSS DR5 \dlas\ visible from the Magellan site, without regard to metallicity, log\nhi , or any other \dla\ property.  Only 2 constraints limited the sample: 1) a lower redshift cutoff of $z_{abs}$ = 2.2 such that the systems could be searched for \htwo , and 2) a magnitude cut of $i\leq$19 such that the sample could be observed in a reasonable amount of time.  
While the initial motivation for this survey was to determine the true covering factor and fraction of \htwo\ in \dlas\ --  the results of which we present in a second paper ~\citep{jorgenson13b} -- we summarize here the major results of this current paper as follows:

\begin{enumerate}

\item{Using spectra primarily taken with the medium-resolution Magellan/MagE spectrograph, we measure the log\nhi , [M/H], [Fe/H], \delvninty , \ewsitwo , $W_{\lambda 1548}$ and $W_{\lambda 2796}$ of a sample of 99 uniformly selected \dlas .}% designed \emph{a priori} to be unbiased.} 

\item{The potential underestimation of metallicity due to possible undetected saturation of spectral features in medium-resolution spectra is determined to be mostly alleviated by applying a flux-based saturation criterion of F$_{min}$/F$_q <$ 0.65. Simulations and tests show that the application of this criterion likely alleviates the worst cases, and any additional underestimation in the metal-line column density will likely be less than $\sim$0.3 dex.}

%\item{The potential underestimation of metallicity due to the `washing-out' of spectral features endured by medium-resolution spectra is determined to be mostly alleviated by applying a flux-based saturation criterion of F$_{min}$/F$_q <$ 0.65. Simulations and tests show that the application of this criterion likely alleviates the worst cases, and any additional underestimation in the metal-line column density will likely be less than $\sim$0.3 dex.}

\item{We determine the redshift evolution in the cosmic mean metallicity over the redshift range, $z$ = [2.2, 4.4] (but note that the majority of \dlas\ fall between $z$ = [2.2, 3.5]), to be $\langle Z \rangle$ = (\fitcosmob )$z -$(1.06$\pm$0.36), an evolution that is somewhat flatter than that found by previous works such as R12, who measure an evolution described by $\langle Z \rangle$ = ($-$0.22$\pm$0.03)$z -$(0.65$\pm$0.09) over z$\sim$[0, 5.5]. Strictly speaking, these slopes are consistent with each other at the 2$\sigma$ level.}

\item{A simple separation of the sample into low and high redshift bins reveals very little evolution in any of the \dla\ parameters, including \delvninty\ and the equivalent width of the Si II $\lambda$1526 transition, \ewsitwo , contrary to the results of ~\cite{ledoux06}.  }

\item{We find a highly significant correlation, $>$9$\sigma$, between \ewsitwo\ and metallicity, similar to that found by ~\cite{pro08}.  \ewsitwo\  and the equivalent width of the high-ion gas tracer, C IV $\lambda$1548\AA , $W_{\lambda 1548}$, are also correlated at the $>$5.5$\sigma$ level.  And similar to the results of ~\cite{murphy07}, we find a correlation between the equivalent width of Mg II $\lambda$2796, $W_{\lambda 2796}$, and metallicity at the level of 2.5$\sigma$.  
}

\end{enumerate}

Initially, the most striking result of the work presented in this paper is the somewhat flatter slope -- consistent with zero -- derived for the evolution of the cosmic mean metallicity. Using this result alone, which is based on the only uniformly selected sample in the literature, we do not see significant evidence for metallicity evolution in \dlas\ in the redshift range $z$ = 2.2 $-$ 4.4. 
The significant evolution found by R12, is heavily weighted by the low \emph{and} high redshift ends of their sample.  These facts lead us to the conclusion that the evolution of cosmic mean metallicity over the redshift range $z$ = 2.2 $-$ 4.4 may %is likely to 
be flatter than that found by R12 and closer to the value presented here. However, we emphasise that our result is strictly consistent with that of R12. 
Logically, the evolution in \dla\ metallicity across cosmic time is expected if models of galaxy formation and evolution and the role played by \dlas\ in that evolution are correct.  In this paradigm, the non-evolution found in the Magellan survey stands out as contradictory.  However, given the limited redshift range and size of the Magellan survey, we caution that, while intriguing, this result is difficult to interpret in the context of the broader cosmological picture without additional unbiased data (or at least data with a well-understood, correctable selection function) including, crucially, both the high ($z >  \sim 4$) and low ($z <  \sim 2$) redshift regimes.

\section*{Acknowledgements}

R. A. J. gratefully acknowledges support from the STFC-funded Galaxy Formation and Evolution programme at the Institute of Astronomy, University of Cambridge and the NSF Astronomy and Astrophysics Postdoctoral Fellowship under award AST-1102683.  M. T. M. thanks the Australian Research Council for a QEII Research Fellowship (DP0877998) and Discovery Project grant (DP130100568).  The authors thank M. Rafelski for sharing data prior to publication.  Australian access to the Magellan Telescopes was supported through the Major National Research Facilities program and National Collaborative Research Infrastructure Strategy of the Australian Federal Government.

\small
\itemindent -0.48cm
\bibliography{regina}

\end{document}